\documentclass[aps,pre,twocolumn,showpacs,showkeys,floatfix,amsfonts,10pt]{revtex4-1}

\usepackage{amsmath}    
\usepackage{verbatim}   
\usepackage{color}      
\usepackage{subfigure}  
\usepackage{hyperref}   
\usepackage{epsfig}
\usepackage{graphicx}
\usepackage{dcolumn}
\usepackage{bm}
\usepackage{amsmath}
\usepackage{graphicx}

\usepackage[all]{xy}
\usepackage[section]{placeins}
 \,
 \,
 \,
 \,

\renewcommand{\vec}[1]{\mbox{\boldmath $#1$}}

\newcommand{\g}{\vec{\Gamma}}

\begin{document}

\title{What does dynamical systems theory teach us about fluids?}
\author{Hadrien Bosetti$^{1,2}$}
\author{Harald A. Posch$^{1}$}

\affiliation{$^1$ Computational Physics Group, Faculty of Physics, University of Vienna, Boltzmanngasse 5, 1090 Wien, Austria}
\affiliation{$^2$ Complex Energy Systems Group, Austrian Institute of Technology, Giefinggasse 6, 1210 Wien, Austria }

\date{\today }

\begin{abstract}
We use molecular dynamics simulations to compute the Lyapunov spectra of many-particle systems resembling simple fluids 
in thermal equilibrium and in non-equilibrium stationary states.  Here we review some of the most interesting  results and point to open questions.

\end{abstract}


\maketitle

\section{Introduction}

 What does dynamical systems theory contribute to our understanding of matter? 
To a large extent,  the royal road to gain an understanding of fluids or solids has been statistical mechanics.
Based on interaction potentials obtained from experiments and quantum mechanical simulations, 
sophisticated perturbation theories are capable of providing a quantitative description  of the static structural properties of such fluids
on the atomistic scale. Computer simulations have provided guidance and invaluable insight to unravel the intricate local structure and
even the non-equilibrium dynamics of ``simple'' liquids including hydrodynamic flows and shock waves. At present,
this program is being extended to ever more complex molecular fluids and/or to systems confined  to particular geometries such as
interfaces and pores, and to fluids out of thermal equilibrium.     

We have raised a related  question, ``What is liquid?'' in 1998 in a similar context \cite{MPH_1998}.
A fluid differs from a solid by the mobility of its particles, and this ability to flow is a collective phenomenon. 
The flow spreads rapidly in phase space, which constitutes the fundamental instability characteristic of a gas or liquid. 
About thirty years ago,  dynamical systems theory provided new tools for the characterization of  the chaotic evolution  in 
a multi-dimensional phase space, which were readily applied to liquids shortly after \cite{PH88,PH_1989}. 
The main idea  is  to follow not only the evolution of a state in phase space but, simultaneously, the evolution of various tiny perturbations 
 applied to that state at an initial time and to measure the growth or decay of such perturbations.  The study of the Lyapunov
 stability, or instability, of a trajectory with respect to small perturbations always present in real physical systems  is hoped
 to provide new and  alternative 
insight into the theoretical foundation  and limitation of dynamical models for dense gases and liquids, of phase transitions involving 
rare events, and of the working of the Second Law of thermodynamics for stationary non-equilibrium systems. 
It is the purpose of this review to assess how far this hope has become true and how much our understanding of fluids has gained from the study of  the Lyapunov instability in phase space.

The structure of a simple fluid is essentially determined by the  steep (e.g. $\propto r^{-12}$)  repulsive part of the pair 
potential, which can be well approximated by a discontinuous hard core. In perturbation theories, this hard potential may be taken  as 
a reference potential with the long-range attractive potential ($\propto - r^{-6}$) acting as a perturbation \cite{Gray}.  
Hard disk fluids in two dimensions, and hard sphere systems in three, are easy to simulate and are paradigms for
simple fluids. The first molecular dynamics simulations for hard sphere systems  were carried out by Alder and Wainwright
\cite{Alder} in 1957, the first computation of Lyapunov spectra for such systems by Dellago, Posch and Hoover 
\cite{DPH_1996} in 1996. 

There exist numerous extensions of the hard-sphere model to include rotational degrees of freedom and various 
molecular geometries \cite{Allen}.  Arguably the simplest, which still preserves spherical symmetry, are spheres with
a rough surface, so-called ''rough hard spheres'' \cite{chapman:1953}. In another approach, fused dumbbell diatomics
are used to model linear molecules \cite{AI_1987,TS_1980}. Both models are used to study the  energy exchange between 
translational and rotational degrees of freedom and, hence, rotation-translation coupling for molecules with 
moderate anisotropy. Other approaches more suitable for larger molecular anisotropies involve spherocylinders
\cite{VB,RS}, ellipsoids \cite{FMM,TAEFK} and even inflexible hard lines \cite{FMag}.  To our knowledge, only for the 
first two schemes, namely for rough hard disks  \cite{BP_2013,Bdiss} and for planar hard dumbbells \cite{MPH_1998,Milano,MPH_chaos},
extensive studies of the Lyapunov spectra have been carried out. We shall come back to this work below.  

There are numerous papers for repulsive soft-potential systems and for Lennard-Jones systems from various authors, in which 
an analysis of the Lyapunov instability has been carried out. We shall make reference to some of this work in the following sections.

The paper is organized as follows. In the next section we introduce global  and local 
(time dependent) Lyapunov exponents and review  some of their properties.  Of  particular interest are the
symmetry properties of the local exponents and of the corresponding perturbation vectors in tangent space. Both 
the familiar orthonormal Gram-Schmidt vectors and the covariant Lyapunov vectors are considered.  In Sec.~\ref{hard_wca} we study planar hard disk systems over a wide range of densities, and 
compare them to analogous fluids interacting with a smooth repulsive potential.  There we also demonstrate that
the largest (absolute) Lyapunov exponents  are generated by perturbations, which are strongly localized in physical space:
only a small cluster  of particles contributes  at any instant of time. This localization persists in the
thermodynamic limit.  Sec.~\ref{hydro} is devoted to another property of the perturbations associated with the 
small exponents, the so-called Lyapunov modes. In a certain sense, the Lyapunov modes are analogous to the 
Goldstone modes of fluctuating hydrodynamics (such as the familiar sound and heat modes). However, it  is
surprisingly difficult to establish a connection between these two different viewpoints. In Sec.~\ref{roto}  
particle systems with translational and
rotational dynamics are considered. Two simple planar model fluids are compared, namely  gases of rough hard disks  and 
of hard-dumbbell molecules. Close similarities, but also surprising differences, are found. Most surprising
is the fast disappearance of the Lyapunov modes and  the breakdown of the
symplectic symmetry of the local Gram-Schmidt exponents for the rough hard disk systems, if the moment of inertia
is increased from zero. In Sec.~\ref{nonequilibrium} we summarize some of the results for stationary systems far from thermal equilibrium, 
for which the Lyapunov spectra have been valuable guides for our understanding.  
For dynamically thermostatted systems
in stationary non-equilibrium states, they provide a direct link  with the Second Law of thermodynamics
due to the presence of a multifractal phase-space distribution. We conclude with a few short remarks
in Sec.~\ref{outlook}.

\section{Lyapunov exponents and perturbation vectors}
\label{general_intro}

Let ${\bf \Gamma}(t)= \{{\bf p},{\bf q}\}$ denote the instantaneous state of a dynamical particle system with a phase space $M$ of dimension $D$.
Here, ${\bf p}$ and ${\bf q}$ stand for the momenta and positions of all the particles. The motion equations are usually written
as a system of first-order differential equations,
\begin{equation}
    \dot{\bf \Gamma} = {\bf F}({\bf \Gamma}),
   \label{motion} 
\end{equation} 
where ${\bf F}$ is a (generally nonlinear) vector-valued function of dimension $D$. The 
formal solution of this equation is written as  ${\bf \Gamma}(t) =  \phi^t ({\bf \Gamma}(0))$, 
where  the map $\phi^t:  {\bf \Gamma} \to {\bf \Gamma}$ defines the flow in $M$, which maps a
phase point ${\bf \Gamma}(0)$ at time zero  to a point ${\bf \Gamma}(t)$ at time $t$.

Next,  we consider an arbitrary infinitesimal perturbation vector $\delta {\bf \Gamma}(t)$, which
depends on the position ${\bf \Gamma}(t)$ in phase space and, hence, implicitly on time.
It evolves according to the linearized equations of motion,
\begin{equation}
    \dot{\delta {\bf \Gamma}} =  {\cal J}({\bf \Gamma})  \delta {\bf \Gamma}.
\label{linearized}
\end{equation}    
This equation  may be formally solved according to
\begin{equation}
\delta{\bf \Gamma}(t) =  D\phi^t \vert_{{\bf \Gamma}(0)}\; \delta{\bf \Gamma}(0),
\label{evolution}
\end{equation}
where $D\phi^t$ defines the flow in tangent space and is represented by a real but
generally non-symmetric $D \times D$ matrix . The  dynamical (or Jacobian) matrix,
\begin{equation}
{\cal J}({\bf \Gamma}) \equiv \frac{ \partial {\bf F}}{ \partial {\bf  \Gamma}},
\label{jacobian}
\end{equation}
determines, whether the perturbation vector $ \delta {\bf \Gamma}(t) $  
has the tendency to grow or shrink at a particular point ${\bf \Gamma}(t)$ in phase space the
system happens to be at time $t$.  Accordingly, the matrix element 
 \begin{equation}
    \delta {\bf \Gamma}^{\dagger}({\bf \Gamma}) {\cal J({\bf \Gamma})}  \delta {\bf \Gamma}({\bf \Gamma})
 \label{lyageneral}  
 \end{equation}
turns out to be positive or negative, respectively. Here, $^\dagger$ means transposition. If, in addition, the perturbation is normalized,
$||\delta {\bf \Gamma}|| = 1$, and points into particular directions in tangent space to be specified below, 
this matrix element turns out to be a local rate for the growth or decay of  $|| \delta {\bf \Gamma}(t) ||$
and will be referred to as a {\em local} Lyapunov exponent $\Lambda({\bf  \Gamma})$ at the phase point
${\bf \Gamma}$.
\subsection{Covariant Lyapunov vectors}
In 1968 Oseledec published his celebrated multiplicative ergodic theorem \cite{Oseledec:1968,Ruelle:1979,Eckmann:1985,Ruelle:1999},
in which he proved that under very general assumptions (which apply to molecular fluids) the tangent space
decomposes into subspaces $E^{(j)}$ with dimension $m_{j}$,
\begin{equation}  
TM({\bf \Gamma}) = E^{(1)}({\bf \Gamma}) \oplus E^{(2)}({\bf \Gamma}) \oplus \cdots \oplus E^{(L)}({\bf \Gamma}),
\label{split}
\end{equation}
for almost all ${\bf \Gamma} \in M$ (with respect to the Lesbegue measure), such that  $\sum_{j=1}^L m_j = D$.
These subspaces evolve according to 
\begin{equation}
 D \phi^t \vert_{{\bf \Gamma}(0)} \; E^{(j)}\left({\bf \Gamma}(0)\right) = E^{(j)}\left( {\bf  \Gamma}(t)\right) , 
\label{defcov}
\end{equation}
and are said to be covariant,  which means that they co-move -- and in particular co-rotate -- with the flow in tangent space. 
In general they are not orthogonal to each other.
If $ {\vec v}\left({\bf \Gamma}(0)\right) \in E^{(j)}\left({\bf \Gamma}(0)\right)$ 
is a vector  in  the subspace $E^{(j)}\left({\bf \Gamma}(0)\right)$, it  evolves according to
\begin{equation}
 D \phi^t \vert_{{\bf \Gamma}(0)} \; \vec v\left({\bf \Gamma}(0)\right)
  = \vec v\left( {\bf  \Gamma}(t)\right)  \in E^{(j)}\left({\bf \Gamma}(t)\right). 
\label{defcovv}
\end{equation} 
The numbers
\begin{equation}
 ^{(\pm)}\lambda^{(j)}   = \lim_{t \rightarrow \pm \infty} \dfrac{1}{\vert t \vert} \, \ln 
         \frac{  \big\|{\vec v} \left({\bf \Gamma}(t)\right) \big\|} { \big\|{\vec v} \left({\bf \Gamma}(0)\right) \big\|} 
\label{covlambda}
\end{equation}
for $j \in \{ 1,2,\dots, L\} $ exist and  are referred to as  the (global) Lyapunov exponents. 
Here, the upper index $(+)$ or $(-)$  indicates, whether the trajectory
is being followed forward or backward  in time, respectively. 
If a  subspace dimension $m_{j}$ is larger than one,  then the respective exponent  is called
 degenerate with multiplicity $m_{j}$. If all exponents are repeated according to their
multiplicity, there are $D$ exponents altogether, which   are commonly ordered according to size, 
\begin{eqnarray}
^{(+)}\lambda_1& \ge & \cdots  \ge ^{(+)}\lambda_D, \label{e1}\\
^{(-)}\lambda_1& \ge & \cdots  \ge ^{(-)}\lambda_D, \label{e2} \\
\nonumber
\end{eqnarray}
where the  subscripts are referred to as Lyapunov index. 
The vetors ${\vec v}^{\ell}$ generating 
$\lambda_{\ell}$ according to 
\begin{equation}
 ^{(\pm)}\lambda_{\ell}   = \lim_{t \rightarrow \pm \infty} \dfrac{1}{\vert t \vert} \, \ln 
         \frac{  \big\|{\vec v}^{\ell} \left({\bf \Gamma}(t)\right) \big\|} { \big\|{\vec v}^{\ell} \left({\bf \Gamma}(0)\right) \big\|}; \;\;\;
         \ell = 1,2,\dots,D,
\label{covlambda1}
\end{equation}
are called covariant Lyapunov vectors. This notation, which treats the tangent space dynamics in terms of a set
of vectors,   is more convenient for algorithmic purposes, although the basic theoretical objects are the covariant subspaces 
$E^{(j)}; \; j = 1,2,\dots,L.$

Degenerate Lyapuov exponents appear, if there exist
 intrinsic continuous symmetries (such as invariance of the Lagrangian with respect to  time and/or 
space translation, giving  rise to energy and/or momentum conservation, respectively).  For particle systems such symmetries almost
always exist. Some consequences  will be  discussed below.

The global exponents for systems evolving forward or backward in time are related according to
\begin{equation}
       ^{(+)}\lambda_{\ell} = - ^{(-)}\lambda_{D+1-\ell}; \: \; \ell \in \{1,2, \dots , D\}. 
\label{fb}
\end{equation}       
The subspaces $E^{(j)}$ and, hence, the covariant Lyapunov vectors
${\vec v}^{\ell}$ generally are not pairwise orthogonal.

If in Eq.~(\ref{covlambda}) the time evolution is not followed from zero to infinity but only over a finite time interval $\tau > 0$, so-called
covariant {\em finite-time dependent} Lyapunov exponents are obtained,
\begin{equation}
 ^{(\pm)}\bar{\Lambda}_{\ell}^{\tau,\mbox{cov}}  =  \dfrac{1}{ \tau} \, \ln \, \big\| \, D\phi^{ \pm \tau}\vert_{\textrm{\smallskip{\g}(0)}} \,
        \; \, {\vec v}^{\ell} \left({\bf \Gamma}(0)\right) \, \big\|.
\label{covftlambda}
\end{equation}
In the limit $\tau \to 0$ so-called {\em local} Lyapunov exponents are generated,
\begin{eqnarray}
 ^{(\pm)}\Lambda_{\ell}^{\mbox{cov}}\left({\bf \Gamma}\right)  & = & \lim_{t \rightarrow \pm 0} \dfrac{1}{\vert t \vert} \, \ln \, \big\| \, \ 
       D\phi^t\vert_{\textrm{\smallskip{\g}}} \, \; \, {\vec v}^{\ell} \left({\bf \Gamma}\right) \, \big\|   \nonumber \\
       & = &   \left[ ^{(\pm)}{\vec v}^{\ell}({\bf \Gamma})  \right]^{\dagger} {\cal J({\bf \Gamma})} \;\; ^{(\pm)}{\vec v}^{\ell}({\bf \Gamma}) 
       \label{covllambda}
\end{eqnarray}
where $\vert\vert{\vec v}^{\ell}({\bf \Gamma})|| = 1$ is required. The second equality has a structure as in
Eq.~(\ref{lyageneral}), applied to the covariant vectors,  and indicates that the local exponents are point functions, which 
only depend on ${\bf  \Gamma}$. The finite-time-dependent exponents may be viewed as  time averages of local exponents over a 
stretch of trajectory during the finite time  $\tau$, the global exponents as  time averages over an infinitely-long trajectory.
For the latter to become dynamical invariants, one requires ergodicity, which we will always assume in the following for
lack of other evidence.  
\subsection{Orthogonal Lyapunov vectors}
Another definition of the Lyapunov exponents, also pioneered by Oseledec  
\cite{Oseledec:1968,Ruelle:1979,Eckmann:1985,Ruelle:1999}, is via the 
real and symmetrical matrices 
\begin{equation}
\lim_{t \to \pm \infty} \left[  D \phi^{t}\vert_{{\bf  \Gamma}(0)}^{\dagger}   D \phi^{t}\vert_{{\bf  \Gamma}(0)} \right]^{\frac{1}{2|t|}},
\label{Oseledec1}
\end{equation}
which exist with probability one (both forward and backward in time). Their (real) eigenvalues involve the global
Lyapunov exponents,
\begin{equation}
  \exp(^{(\pm)}\lambda_1) \ge \cdots  \ge  \exp(^{(\pm)}\lambda_D),
\end{equation}  
where, as before,  degenerate  eigenvalues are repeated according to their multiplicities.
For non-degenerate and degenerate eigenvalues the $>$ and $=$ signs apply, respectively.
The corresponding eigenspaces, $U_{\pm}^{(j)}; \; j  = 1,\dots, L$, are pairwise orthogonal and provide two 
additional decompositions of the tangent space at almost every point in phase space, one forward $(+)$ and one backward $(-)$ in time:
\begin{equation}  
TM({\bf \Gamma}) = U_{\pm}^{(1)}({\bf \Gamma}) \oplus U_{\pm}^{(2)}({\bf \Gamma}) \oplus \cdots \oplus U_{\pm}^{(L)}({\bf \Gamma}).
\label{splitalt}
\end{equation}
In each case, the dimensions $m_j$ of the Oseledec subspaces $U_{\pm}^{(j)}; \; j = 1,\dots,L, $ have a sum equal  to the phase-space dimension $D$:
$\sum_{j=1}^L m_j = D.$
These subspaces are  not  covariant.

The classical algorithm for the computation of (global) Lyapunov exponents  \cite{Benettin,Shimada,Wolf}  carefully keeps track of
all $d$-dimensional infinitesimal volume elements $\delta {V}^{(d)}$ which (almost always) evolve according to
\begin{equation}
 \delta {V}^{(d)}(t)  \approx    \delta {V}^{(d)}(0)  \exp \left( \sum_{\ell  = 1}^d \lambda_{\ell} t  \right).
 \label{volume}
\end{equation} 
This is algorithmically achieved with the help of a set of perturbation vectors $^{(\pm)}{\vec g}_{\ell}({\bf \Gamma}); \;\; \ell = 1,\dots,D$, which are either  periodically re-orthonormalized 
with a Gram-Schmidt (GS) procedure (equivalent to a QR-decomposition) \cite{recipes}, or are  continuously  constrained to stay
orthonormal with a triangular matrix of Lagrange multipliers \cite{HP1987,Goldhirsch,PH88,PH04,BPDH}. For historical reasons we 
refer to them as 
GS-vectors. They have been shown to be the spanning vectors for the Oseledec subspaces   $U_{\pm}^{(j)}$ \cite{Ershov}. 
Accordingly, they are  orthonormal but not covariant.

Analogous to Eqs.~(\ref{covftlambda}) and~(\ref{covllambda}),  one may  define finite-time-dependent GS Lyapunov exponents 
$ ^{(\pm)}\bar{\Lambda}_{\ell}^{\tau,\mbox{GS}}$  and local GS exponents $ ^{(\pm)}\Lambda_{\ell}^{\mbox{GS}}\left({\bf \Gamma}\right)$,
where the latter are again point functions,
\begin{equation}
 ^{(\pm)}\Lambda_{\ell}^{\mbox{GS}}\left({\bf \Gamma}\right)    =  [ ^{(\pm)}{\vec g}^{\ell}({\bf \Gamma})  ]^{\dagger}\; {\cal J({\bf \Gamma})}  
 \;\;^{(\pm)}{\vec g}^{\ell}({\bf \Gamma}). 
\end{equation}
As before, the finite-time and global GS exponents are time averages  of the respective local exponents over a finite  
respective infinite trajectory \cite{Ershov,BPDH}.

From all GS vectors only  $^{(\pm)}{\vec g}_{1}({\bf \Gamma})$, which is associated with the maximum global
exponent, evolves freely without constraints which might affect its  orientation in tangent space. Therefore it agrees with 
$^{(\pm)}{\vec v}_{1}({\bf \Gamma})$ and is also covariant. Also the corresponding local exponents agree for $\ell = 1$,
but generally not for other $\ell$. However,  the local covariant exponents may be computed from the local 
GS exponents, and vice versa, if the angles between them are known \cite{BPDH}.

As an illustration of the relation between covariant Lyapunov vectors and orthonormal GS vectors, we consider a
simple two-dimensional example,  the H\'enon map \cite{Henon}, 
\begin{eqnarray}
x_{n+1} &=& a - x_n^2 + b \, y_n
\nonumber
\enspace , \\
y_{n+1} &=& x_n \enspace ,
\nonumber
\end{eqnarray}
with $a=1.4$ and $b=0.3$.  
\begin{figure}[t]
\includegraphics[width=0.4\textwidth]{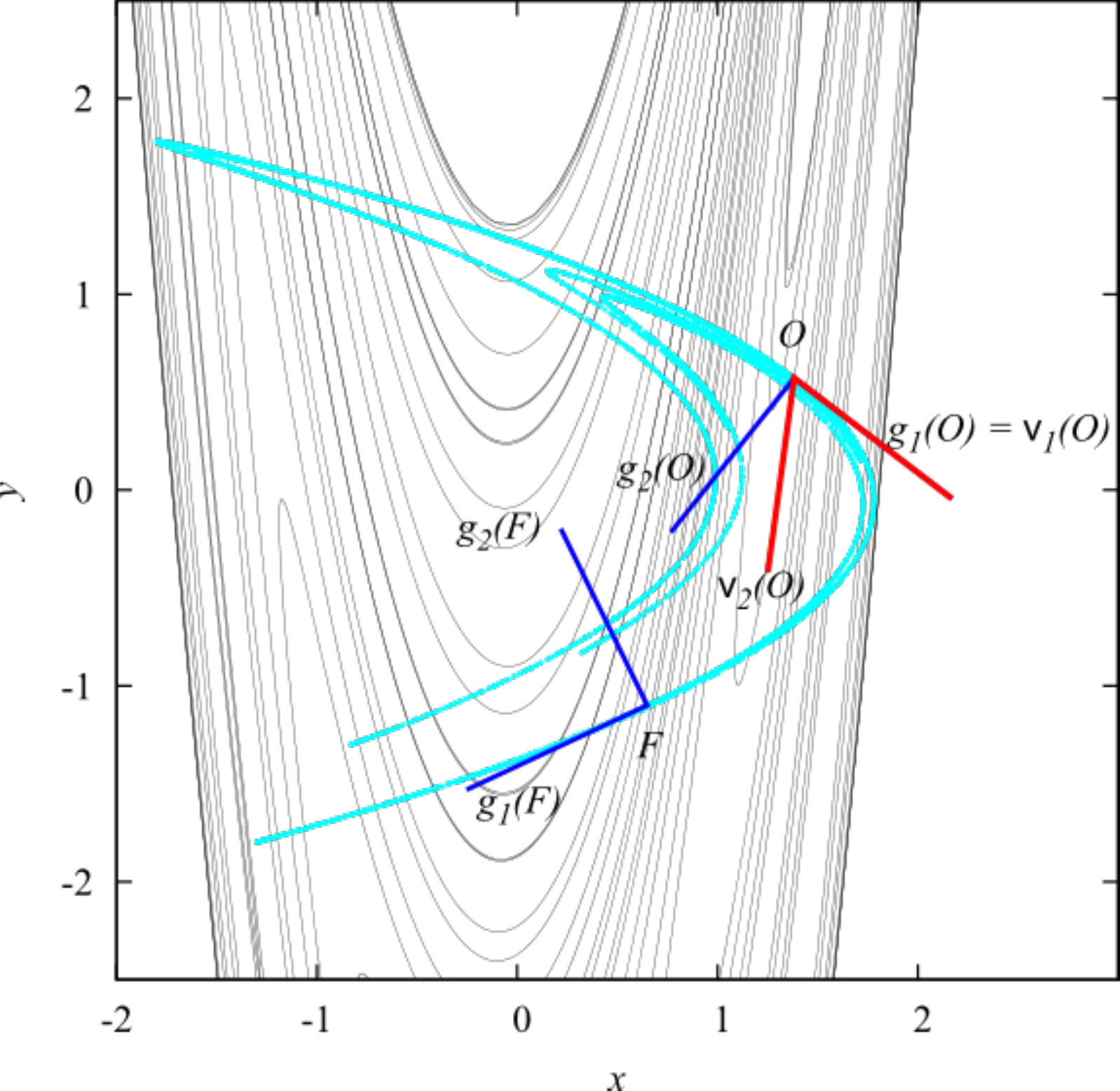}
\caption{(Color online) Covariant and GS vectors for a point $O$ on the H\'enon attractor (light-blue). The black line is a finite-time approximation of the (global) stable manifold.}
\label{henon_map}  
\end{figure}
The light-blue line in Fig.~\ref{henon_map} represents the H\'enon attractor, which is known to coincide with its unstable manifold. 
The black line is a finite-time approximation of its stable manifold. Let $O$ denote a point on the attractor.
The GS vectors at this point are indicated by ${\vec g}_1(O)$ and ${\vec g}_2(O)$, where the former is also covariant and 
identical to ${\vec v}_1(O)$ (parallel to the unstable manifold). As required, the covariant vector ${\vec v}_2(O)$ is tangent to the
stable manifold  (which was determined by a different method).  For the computation of the covariant vectors at $O$, 
it is necessary to follow and store the reference trajectory and the GS vectors sufficiently far into the future up to a point $F$
in our example. In Fig.~\ref{henon_map} the  GS vectors at this point are denoted by ${\vec g}_1(F)$ and  ${\vec g}_2(F)$. Applying an algorithm
by Ginelli {et al.} \cite{Ginelli}, to which we shall return below, a backward iteration to the original phase point $O$ yields the covariant vectors,
${\vec v}_2(O)$ in particular.

\subsection{Symmetry properties of global and local exponents}

For ergodic systems the global exponents are averages of the local exponents over the natural measure
in phase space and, according to the multiplicative ergodic theorem for chaotic systems, they do not depend on 
the metric and the norm one applies to the tangent vectors.  Also the choice of the coordinate system 
(Cartesian or polar, for example) does not matter.
For practical reasons the Euclidian (or L2) norm will be used throughout in the following. It also
does not matter, whether covariant or Gram-Schmidt vectors are used for the computation.
The global exponents  are truly dynamical invariants. 

This is not the case for the local exponents. They depend on the norm and on the coordinate system.
And they generally depend on the set of basis vectors, covariant or GS,  as was mentioned before. Furthermore, some
symmetry properties of the equations of motion are reflected quite differently  by the two local representations.
\begin{itemize}
\item  Local Gram-Schmidt exponents: During the construction of the GS vectors, the changes of the 
differential volume elements $\delta {V}^{(d)}$  following Eq.~(\ref{volume}) are  used to compute the local exponents.  
If the total phase volume is conserved as is the case for time-independent Hamiltonian systems, the following sum rule 
holds for almost all ${\bf \Gamma}$:
\begin{equation}
            \sum_{\ell = 1}^D \Lambda_{\ell}^{\mbox{GS}}({\bf \Gamma}) = 0.
\label{sum_rule}
\end{equation}
In this symplectic case we can even say more. For each positive local GS exponent there exists a 
negative local GS exponent such that their pair sum vanishes \cite{Meyer}: 
\begin{equation}
^{(\pm)}\Lambda_{\ell}^{\mbox{GS}}({\bf \Gamma}) =  -^{(\pm)}\Lambda_{D + 1 -\ell }^{\mbox{GS}}({\bf \Gamma}),\;\; \ell = 1,\dots,D. \label{gsfwd} 
\end{equation}
As indicated, such a symplectic local pairing symmetry is found both forward and backward in time.
Non-symplectic systems do not display that  symmetry.
On the other hand, the re-orthonormalization process tampers with the orientation and rotation of the GS vectors
and destroys all consequences  of the  time-reversal invariance property of the original motion equations.  
 
 \item Local covariant exponents: During their construction \cite{Ginelli}, only the norm of the  covariant perturbation vectors 
needs to be periodically adjusted for practical reasons,  but the angles between them remain unchanged.  This process effectively destroys 
all information concerning the $d$-dimensional volume elements. Thus, no symmetries 
analogous to Eq.~(\ref{gsfwd})  exist.  Instead, the re-normalized covariant vectors faithfully 
preserve the time-reversal symmetry of the equations of motion, which is  reflected by
\begin{equation}   
 ^{(\mp)}\Lambda_{\ell}^{\mbox{cov}}({\bf \Gamma})  = -^{(\pm)}\Lambda_{D+1 -\ell}^{\mbox{cov}}({\bf \Gamma}), \; \;  \ell = 1,\dots,D,
 \label{local_symmetry}
\end{equation}
regardless, whether the system is symplectic or not. This means that an expanding co-moving direction is
converted into a collapsing co-moving direction by an application of the time-reversal operation.
\end{itemize}

The set of all global Lyapunov exponents, ordered according to size, is referred to as the Lyapunov spectrum. For stationary Hamiltonian systems in thermal equilibrium the (global) conjugate pairing rule holds,
 \begin{equation}
^{(\pm)}\lambda_{\ell} =  -^{(\pm)}\lambda_{D + 1 -\ell }, \label{cpr} 
\nonumber
\end{equation}
which is a consequence of Eq.~(\ref{gsfwd}) and of the fact that the global exponents are the phase-space averages of these quantities.
In such a case only the first half of the spectrum containing the positive exponents needs to be computed. In the following we shall
refer to this half as the positive branch of the spectrum.

The subspaces spanned by the covariant (or GS) vectors associated with the strictly 
positive/negative  global exponents are  known as the unstable/stable manifolds.
Both are covariant and are linked by the time-reversal transformation, which converts one into the other.

\subsection{Numerical considerations}

The computation of the Gram-Schmidt exponents is commonly carried out with the algorithm of
Benettin {\em et al.} \cite{Benettin,Wolf} and Shimada {\em et al.} \cite{Shimada}. Based on this classical approach, a 
reasonably efficient algorithm for the computation of covariant exponents has been recently developed by  Ginelli {\em et al.}
\cite{Ginelli,Ginelli_2013}. Some computational details for this method may also be 
found in Refs.~\cite{BPDH,BP_2010,BP_2013}. 
An alternative approach is due to Wolfe and Samelson  \cite{Wolfe}, which was subsequently applied 
to Hamiltonian systems with many degrees of freedom \cite{Romero}.

The considerations of the previous section are for time-continuous systems based on the differential equations (\ref{motion}). They may be 
readily extended to maps such as systems of hard spheres, for which a pre-collision state of two colliding particles is 
instantaneously mapped to an after-collision state. Between collisions the particles move smoothly.   
With the linearized  collision map the time evolution of the perturbation vectors 
in tangent space may be constructed  \cite{DPH_1996}. 

In numerical experiments of stationary systems, the initial orientation of the perturbation vectors is arbitrary.
There exists a transient time, during which the perturbation vectors
converge to their proper orientation on the attractor. All symmetry relations mentioned above refer to this
well-converged state and exclude transient conditions  \cite{WHH}.   

For the computation of the full set of exponents, the reference trajectory and $D$ perturbation vectors  
(each of dimension $D$) have to be followed in time, which  requires $D(D+1)$ equations to be integrated. Present computer 
technology limits the number of particles to about $10^4$ for time-continuous systems, and to   $10^5$ for hard-body systems.. 
In all applications below appropriate reduced units will be used.

To ease the notation, we shall in the following omit to  indicate the forward-direction in time by $(+)$, if there is no ambiguity.

\section{Two-dimensional hard-disk and WCA fluids in equilibrium}
\label{hard_wca}

\begin{figure}
\includegraphics[width=0.35\textwidth,angle=-90]{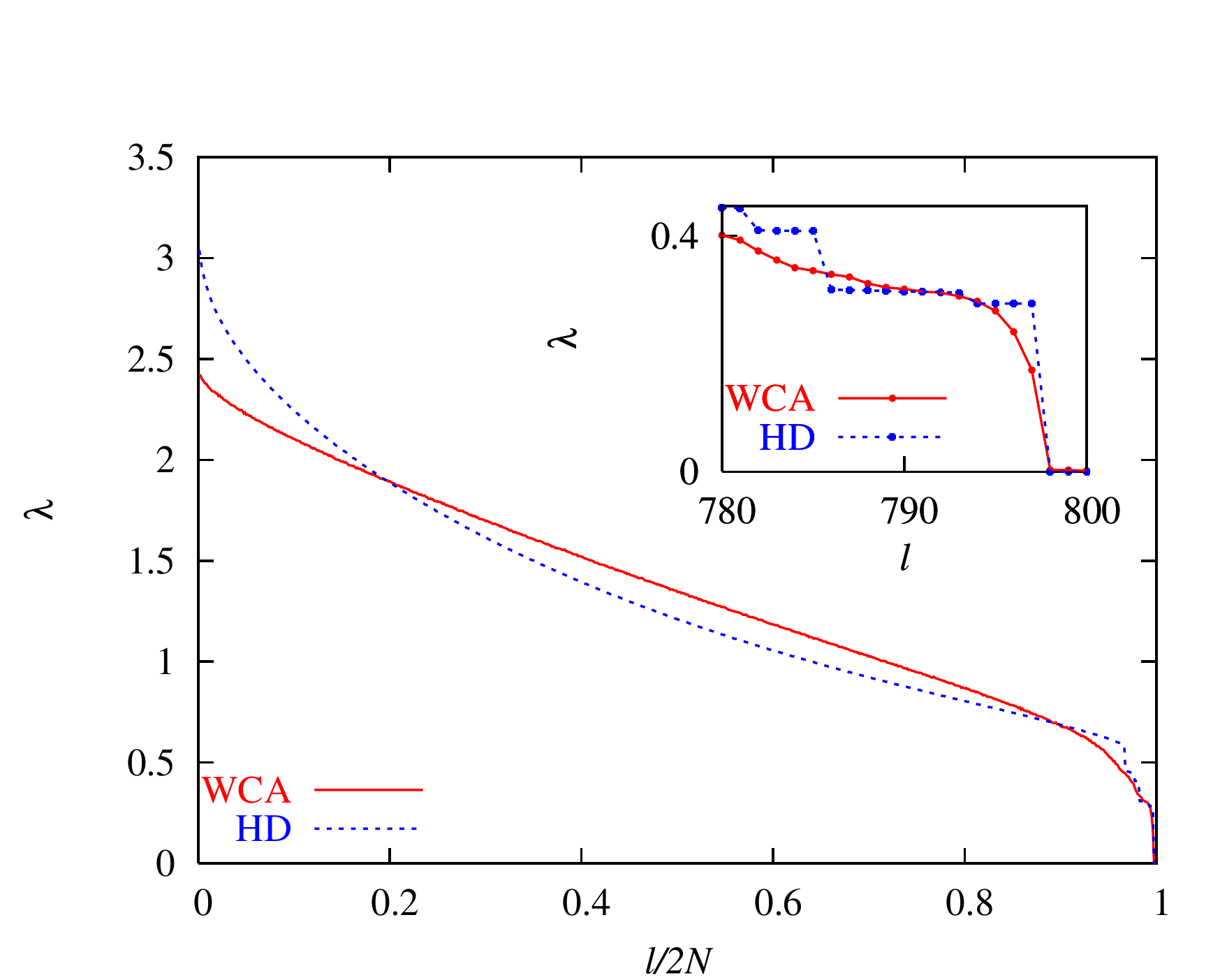}
 \caption{(Global) Lyapunov spectra of a planar  hard-disk gas  and of a planar  WCA fluid with the same number of particles, 
 density and temperature. For details we refer to the main text. Only the positive branches of the spectra are shown.
 Although the spectra consist of discrete points for integer values of the index $l$, smooth lines and 
 reduced indexes $l/2N$ are used for clarity in the main panel. In the inset a 
  magnified view of the small-exponents regime is shown, for which $l$ is not normalized.
  The figure is taken from Ref.~\cite{FP_2005}.
  }
   \label{comparison} 
\end{figure}
We are now in the position to apply this formalism to various models for simple fluids \cite{FP_2005}. First we consider
two moderately dense planar gases, namely  a system of elastic smooth hard disks with a pair potential
\begin{equation}    
                                   \phi_{HD} = \left\{\begin{array}{ll} \infty, &\qquad r \leq \sigma, \\
                                  0,&\qquad r>\sigma, \end{array}\right.
\nonumber
\end{equation} 
and a  two-dimensional Weeks-Chandler-Anderson (WCA) gas interacting with a repulsive soft  potential \cite{WCA1,WCA2}
\begin{equation}
\phi_{WCA}=\left\{\begin{array}{ll}4\epsilon\left[\left(\frac{\sigma}{r}\right)^{12}-
                                       \left(\frac{\sigma}{r}\right)^6\right]+\epsilon,
                                       &\qquad  
                                       r \leq 2^{1/6}\sigma, \\ 
                                      0,&\qquad r>2^{1/6}\sigma. \end{array}\right. 
\nonumber
\end{equation}
In Fig.~\ref{comparison} the positive branches of their (global) Lyapunov spectra are compared.
Both gases consist of $N = 400$ particles  in a square box with side length $L$ and periodic boundaries, 
and have the same density,  $\rho = N/L^2 = 0.4$, and temperature, $T = 1$.  Although, as expected, for low densities 
the Lyapunov spectra are rather insensitive to the interaction potential, the comparison of
Fig.~\ref{comparison} for moderately dense gases already reveals a rather strong sensitivity. In particular,
the maximum exponent $\lambda_1$ is much lower for the WCA fluid, which means that deterministic chaos is 
significantly reduced in systems with smooth interaction potentials.  This difference becomes even more pronounced 
for larger densities, as will be shown below.

\begin{figure}
\includegraphics[width=0.3\textwidth]{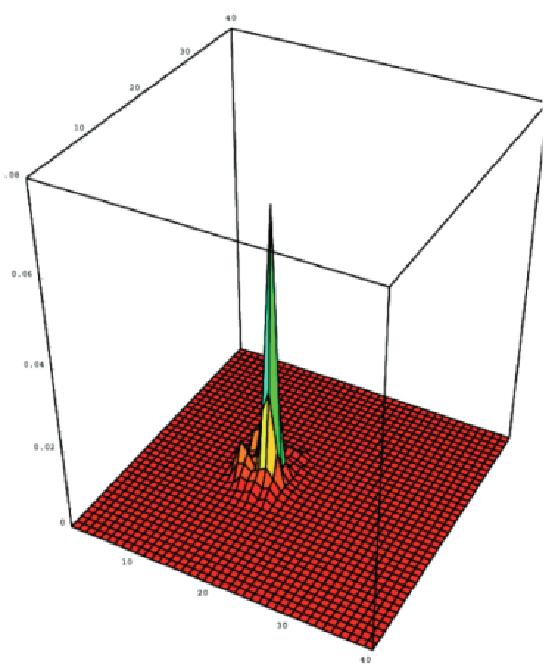}
\caption{Localization of the perturbation vector ${\vec g}^1$ in physical space (the red square) for a 
planar gas of 102,400 soft repulsive disks \cite{HBP_1998}.  The quantity $\mu^{1}$, measuring the 
contribution of individual particles to the perturbation vector associated with the maximum GS exponent, is 
plotted in the vertical direction at the position of the particles. }
\label{local_surface}
\end{figure}
The maximum (minimum) Lyapunov exponent denotes the rate of fastest perturbation growth (decay) 
and is expected to be dominated by the fastest dynamical events such as a locally enhanced collision frequency.
To prove this expectation one may ask how individual particles contribute to the growth of 
the perturbations  determined by  ${\vec v}^{\ell}$  or ${\vec g}^{\ell}$.   
Writing the {\em normalized} perturbation vectors in terms of the position and momentum perturbations of all
particles, ${\vec v}^{\ell} =  \{\delta {\bf q}_i^{(\ell)}, \delta {\bf p}_i^{(\ell)}; i=1,\dots, N  \}$, the quantity
\begin{equation}
\mu_i^{(\ell)} \equiv  \left( \delta {\bf q}_i^{(\ell)}\right)^2 + \left( \delta {\bf p}_i^{(\ell)}\right)^2
\label{gamma}
\end{equation}
is positive,  bounded,  and obeys the sum rule $\sum_{i=1}^N \mu_i^{(\ell)} = 1$ for any $\ell$. 
It may be interpreted as a measure  for the activity of particle $i$ contributing to the perturbation in question. 
Equivalent relations apply for ${\vec g}^{\ell}$.
If  $\mu_i^{(1)}$ is plotted - vertical to the simulation plane - at the position of particle $i$,
surfaces such as in Fig.~\ref{local_surface} are obtained.  They are strongly peaked in  a small domain of the
physical space indicating strong localization of ${\vec v}^{1} \equiv {\vec g}^1$. It means that only a small 
fraction of all particles contributes to the
perturbation growth at any instant of time. This property is very robust and persists in the thermodynamic limit
in the sense that the fraction of particles, which contribute significantly to the formation of ${\vec v}^{1} $
varies as a negative power of $N$ \cite{Milano,FHPH_2004,FP_2005}.
For example, Fig.~\ref{local_surface} is obtained for a system of 102,400 (!) smooth disks interacting with a 
pair potential similar to the WCA potential \cite{HBP_1998}.  

This dynamical localization of the perturbation vectors associated with the large Lyapunov exponents is a very general
phenomenon and has been also observed for one-dimensional  models of space-time chaos \cite{Pikovsky_1} and
Hamiltonian lattices \cite{Pikovsky_2}.

\begin{figure}[t]
\centering
\includegraphics[angle=-90,width=0.47\textwidth]{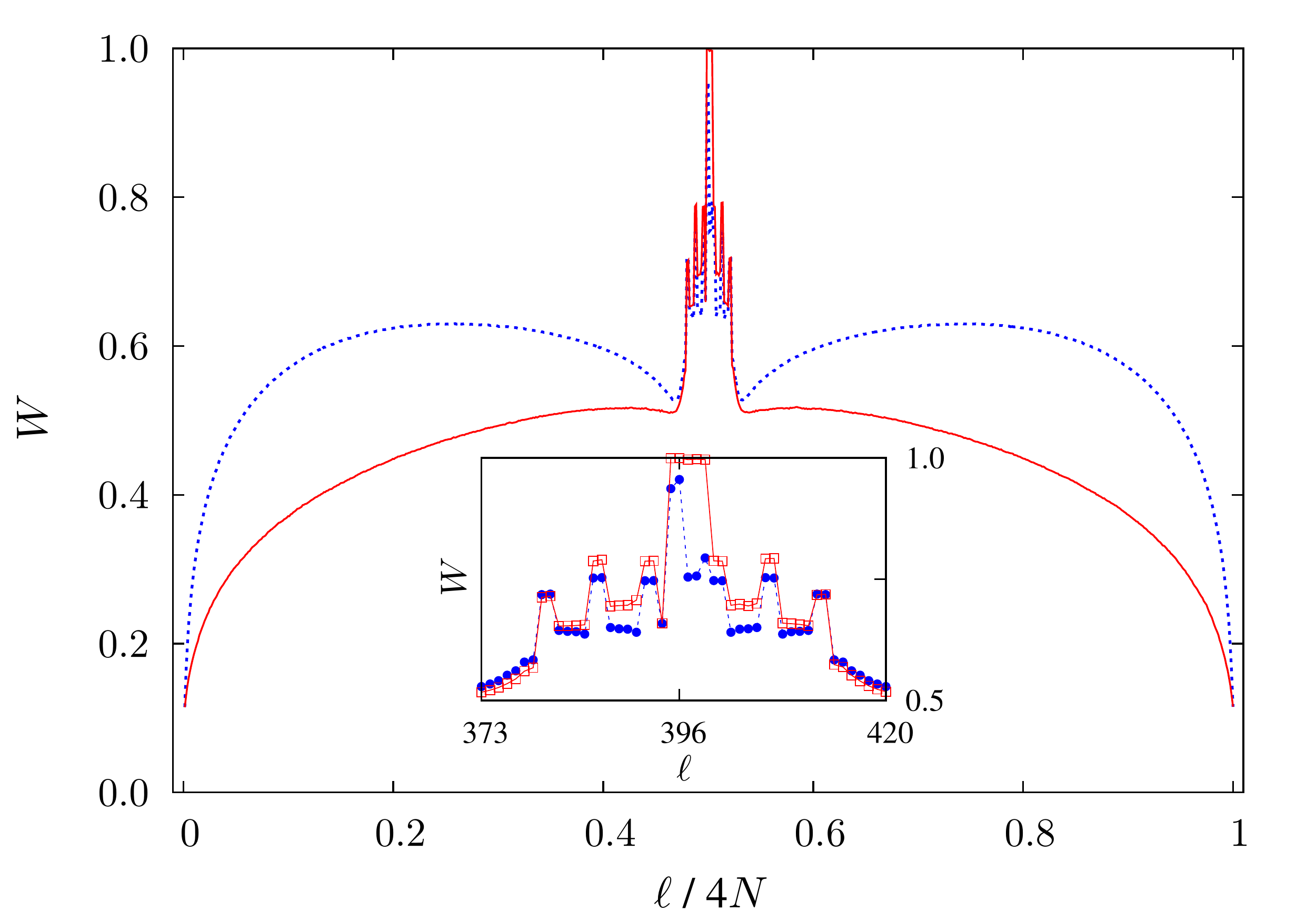}
\caption{Localization spectra $W$ for the complete set of Gram-Schmidt vectors (blue)  
and covariant vectors (red) for 198 hard disks in a periodic box with an aspect ratio 2/11. 
The density $\rho = 0.7$ and the temperature $T = 1$.
Reduced indices $\ell/4N$ are used on the abscissa of the main panel. The inset shows a magnification
of the central part of the spectra dominated by Lyapunov modes. 
 From Ref.~\cite{BP_2010}.}
\label{W}
\end{figure}

For $\ell > 1$, the localization for ${\vec v}^{\ell}$ differs from that of   ${\vec g}^{\ell}$. This may be seen
by using a localization measure due to Taniguchi and Morriss  \cite{TM_2003a,TM_2003b},
\begin{equation}
W = \frac{1}{N} \exp\langle S \rangle; \;\;\;\; S({\bf \Gamma}(t)) = -\sum_{i=1}^N \mu_i^{(\ell)} \ln \mu_i^{(\ell)} ,
\nonumber
\end{equation}
which is bounded according to $1/N \le W \le 1$. The lower and upper bounds correspond to complete localization
and delocalization, respectively. $S$ is the Shannon entropy for the `measure' defined in Eq.~(\ref{gamma}), and
$\langle \dots \rangle$ denotes a time average.
The localization spectra for the two sets of perturbation vectors are shown in
Fig.~\ref{W} for a rather dense system of hard disks  in two dimensions. One may infer from the figure that 
localization is much stronger for the covariant vectors (red line) whose orientations in tangent space 
are only determined by the tangent flow and are not constrained by periodic re-orthogonalization steps
as is the case for the Gram-Schmidt vectors (blue line). One further observes that the localization of the
covariant vectors becomes less and less pronounced the more the regime of coherent Lyapunov modes, located in the center 
of the spectrum, is approached.

Next we turn our attention to the maximum Lyapunov exponent $\lambda_1$ and to the Kolmogorov-Sinai entropy.
Both quantities are indicators of dynamical chaos. The KS-entropy is the rate with which information about an initial state is generated, 
if a finite-precision measurement at some later time is retraced backward along the stable manifold. According to 
Pesin's theorem  \cite{Pesin}  it is equal to the sum of the positive Lyapunov exponents, $h_{KS} = \sum_{\lambda_{\ell} > 0} \lambda_{\ell}$.
It  also determines the  characteristic time for phase space mixing  according to \cite{Arnold,Zaslav,DP_relax}  
$\tau^{mix} = 1/h_{KS}$.

\begin{figure}[t]
 \centering{\includegraphics[width=0.3\textwidth,angle=-90]{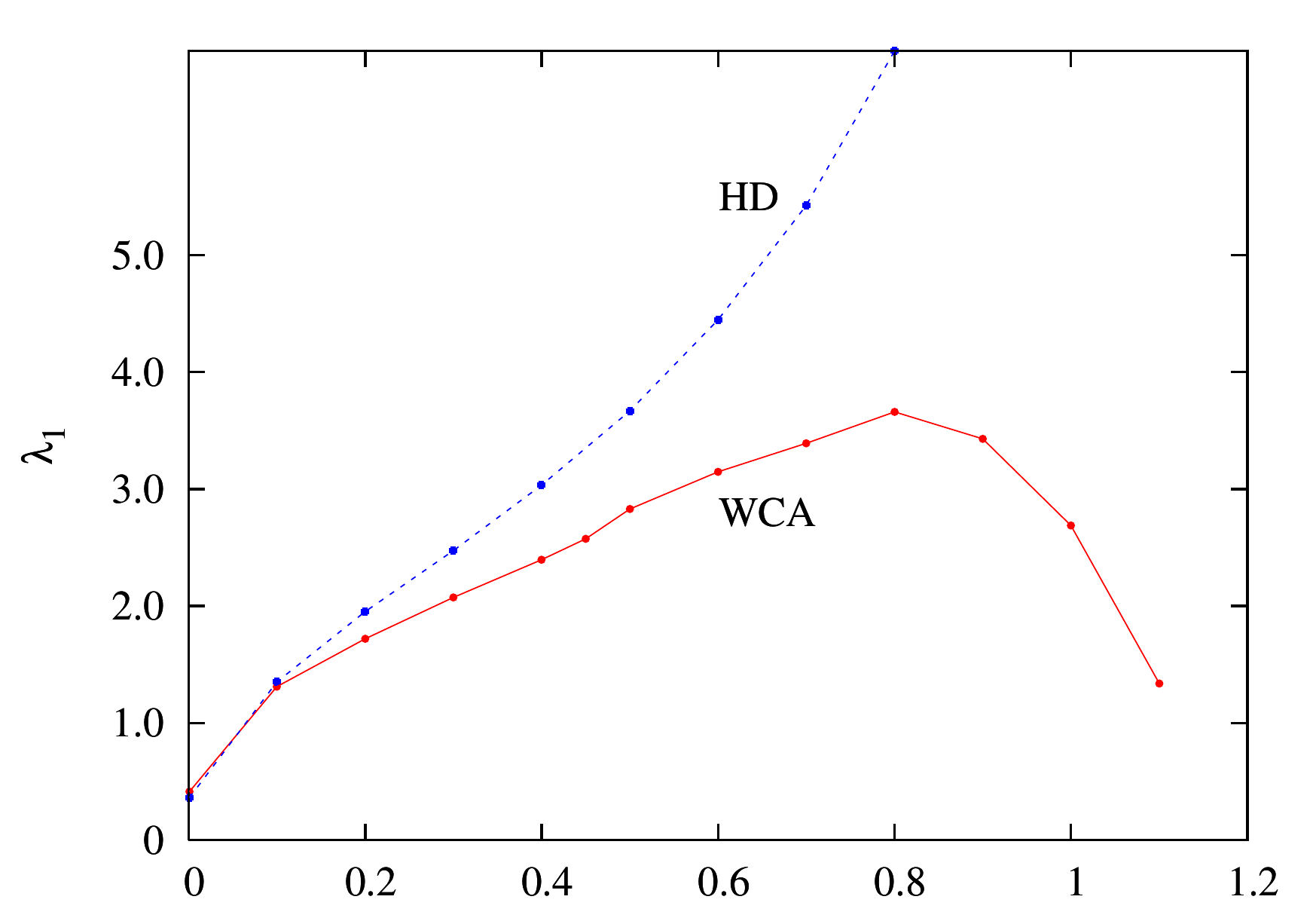}}
\vfill
    {\includegraphics[width=0.3\textwidth,angle=-90]{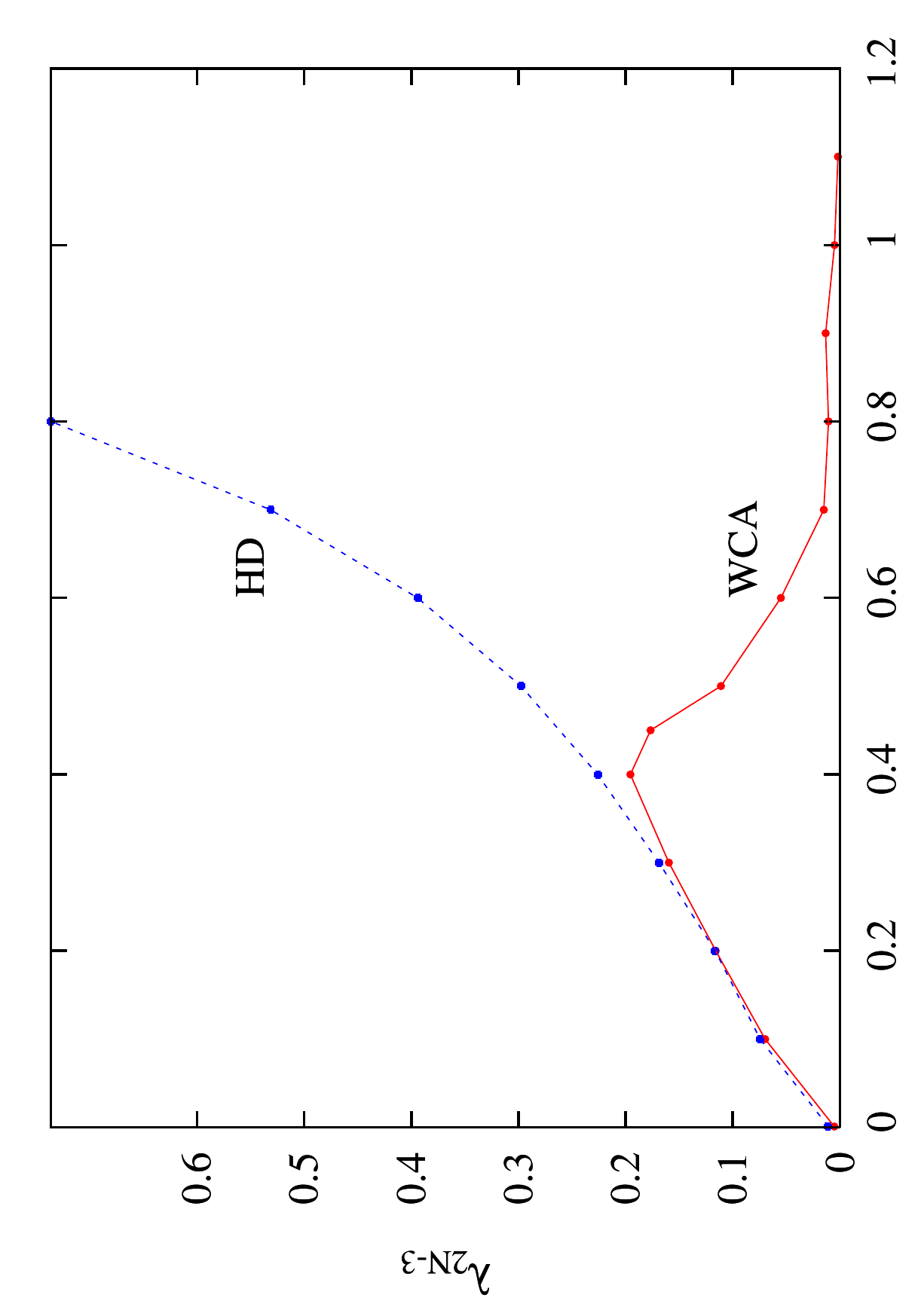}}
\vfill
    {\includegraphics[width=0.3\textwidth,angle=-90]{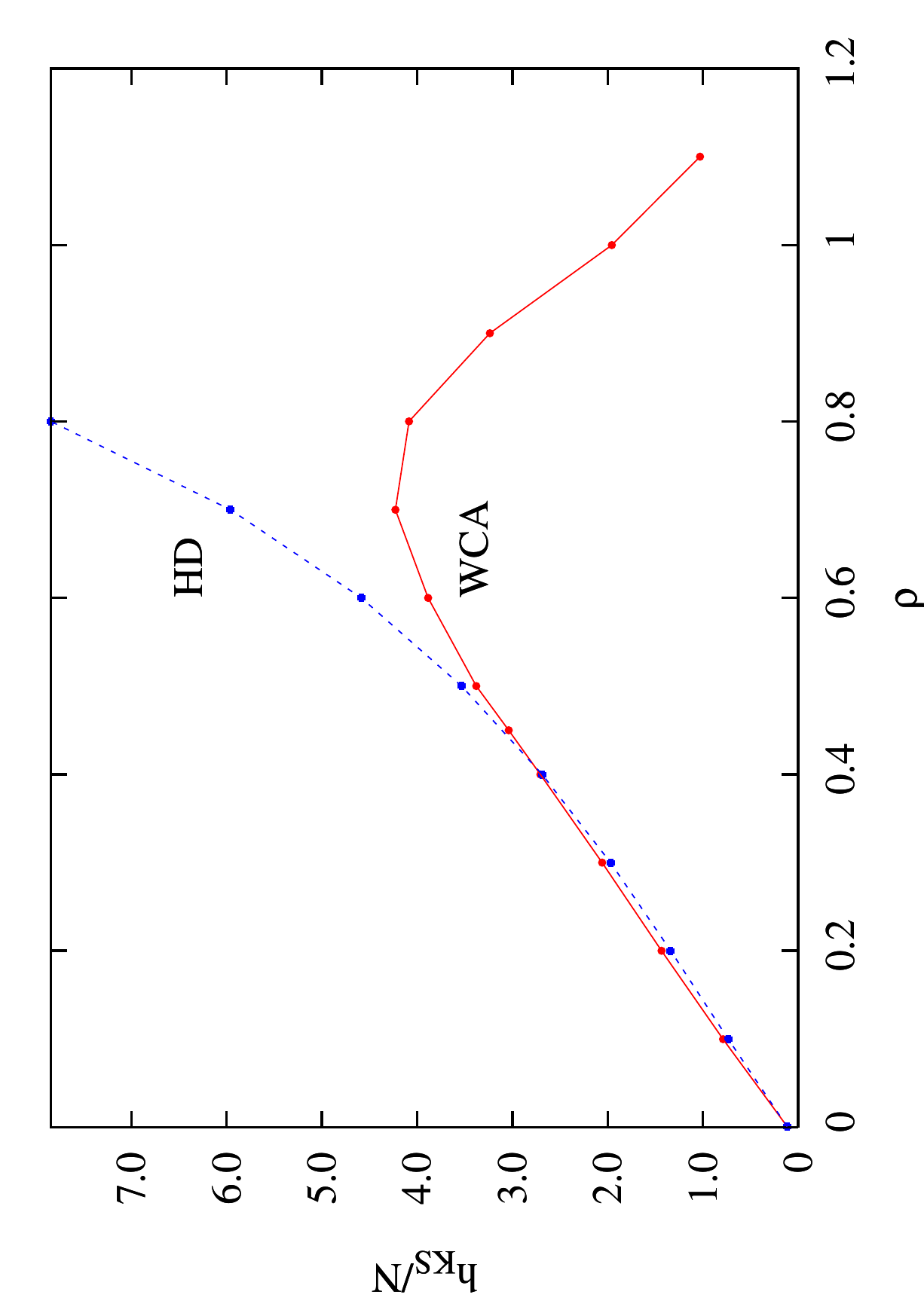}}
\vspace{3mm}
\caption{Isothermal density dependence of the maximum Lyapunov exponent, $\lambda_1$ (top), of 
the smallest positive exponent, $\lambda_{2N-3}$ (middle), and of the Kolmogorov-Sinai entropy 
per particle, $h_{KS}/N$ (bottom), for hard and  soft-disk systems.
From Ref.~\cite{FP_2005}.}
\label{dens}
\end{figure}

In Fig.~\ref{dens} we compare the isothermal density dependence of the maximum Lyapunov exponent $\lambda_1$ (top panel), of the 
smallest positive exponent  $\lambda_{2N-3}$ (middle panel), and of the Kolmogorov-Sinai (KS) entropy per particle $h_{KS}/N$ 
(bottom panel) for planar WCA and HD fluids. Both systems contain $N = 375$ particles  in a periodic box  with aspect ratio 0.6 and with 
a temperature  $T = 1$. As expected, these quantities for the two fluids agree at small densities, but differ significantly 
for large $\rho$.  

For  hard disks, van Zon and van Beijeren  \cite{Zon} and de Wijn \cite{Wijn} used kinetic theory to obtain expressions 
for $\lambda_1$ and $h_{KS}/N$ to leading orders of $\rho$.  They agree very well with  computer simulations of 
Dellago {\em et al.}  \cite{BDPD,ZBD}. The regime of larger densities, however,  is only qualitatively understood. $\lambda_1^{HD}$ and $h_{KS}^{HD}/N$
increase monotonically due to the increase of the collision frequency $\nu$. The (first-order) fluid-solid transition 
shows up as a step between the freezing point of the fluid ( $\rho_f^{HD}= 0.88 $ ) 
and the melting point of the solid ($\rho_m^{HD}=0.91$ \cite{Stillinger}) (not shown in the figure). These steps  disappear, if instead of the density the collision frequency is plotted  on the abscissa. $\lambda_1^{HD}$ and $h_{KS}^{HD}/N$ diverge at the close-packed density due to the divergence of $\nu$.

For the WCA fluid, both the maximum exponent and the Kolmogorov-Sinai entropy have a maximum as a function of density, and become very small when the  density of the freezing point  is approached, which happens for $\rho_f^{WCA} = 0.89 $ \cite{Tox}. 
This behavior is not too surprising in view of the 
fact that a harmonic solid is  not even chaotic and $\lambda_1$ vanishes.  The maximum of $\lambda_1^{WCA}$ occurs 
at a density of about $0.9 \rho_f^{WCA}$ and confirms earlier results for Lennard-Jones fluids \cite{PHH_1990}. 
At this density chaos is most pronounced possibly due to the onset of cooperative dynamics characteristic of phase transitions. 
At about the same density, mixing becomes most effective as  is demonstrated by the maximum of the KS-entropy. The latter is related 
to the mixing time  in phase space according to $\tau^{mix} = 1 / h_{KS}$ \cite{Arnold,Zaslav}.

\begin{figure}[t]
\centering
\includegraphics[width=0.47\textwidth]{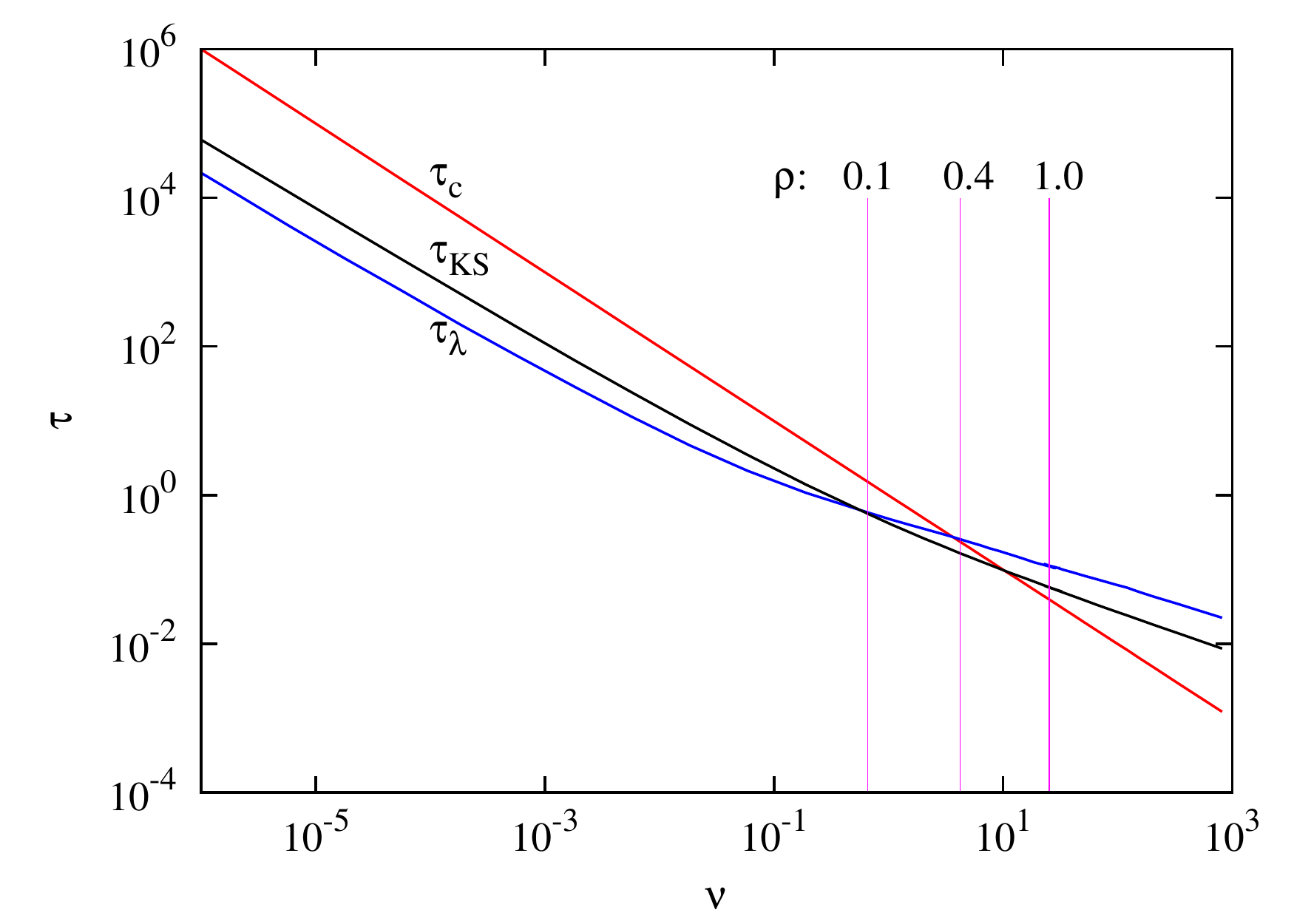}
\caption{(Color online) Comparison of the Lyapunov time $\tau_{\lambda} = 1/\lambda_1$, the time  $ \tau_{KS} \equiv  N/h_{KS}$,  and the collision time 
$\tau_c = 1/\nu$,  as a function of the collision frequency $\nu$,  for a system of 108 hard spheres in a cubic box with periodic boundaries. 
The vertical lines indicate collision frequencies corresponding to the densities $\rho=0.1, 0.4,$ and $1.0$.
}
\label{tau}
\end{figure}

It  is interesting to compare the Lyapunov time $\tau_{\lambda} = 1/\lambda_1$, which is a measure for  the system to ``forget''  its past, 
with the (independently-measured) time between collisions of a particle, $\tau_c = 1/\nu$. In Fig.~\ref{tau} such a comparison is shown for  a 
three-dimensional system of 108 hard spheres in a cubic box with periodic boundary conditions  \cite{DP_relax,DP_3d}.
Also included is the behavior of $\tau_{KS} \equiv N / h_{KS}$. For small densities,  we have $\tau_{\lambda} \ll \tau_c$,
and subsequent collisions are uncorrelated. This provides the basis  for the validity of lowest-order kinetic theory (disregarding
correlated collisions). For densities $\rho > 0.4$ the Lyapunov time is progressively larger than the collision time  and higher-order corrections such as ring collisions
become important. Also the lines for $\tau_{\lambda}$ and $ \tau_{KS}$  cross even before, at a density 0.1. This is a
consequence of the  shape change of the Lyapunov spectrum with density  \cite{DP_3d}: the small positive exponents grow faster with $\rho$ than the large exponents. It also follows that neither the Lyapunov time nor the mixing time determine the 
correlation decay   of - say - the particle velocities. For lower densities, for which ring collisions are not important, the decay of the
velocity autocorrelation function  $z(t)$ is strictly dominated  by the collision time. In Ref.~\cite{DP_relax}
we also demonstrate that for hard-particle systems the time $\tau_{\lambda}$ does not provide an upper bound for the time for which 
correlation functions  may be reliably computed.

\begin{figure}[h]
\centering
\includegraphics[width=0.47\textwidth]{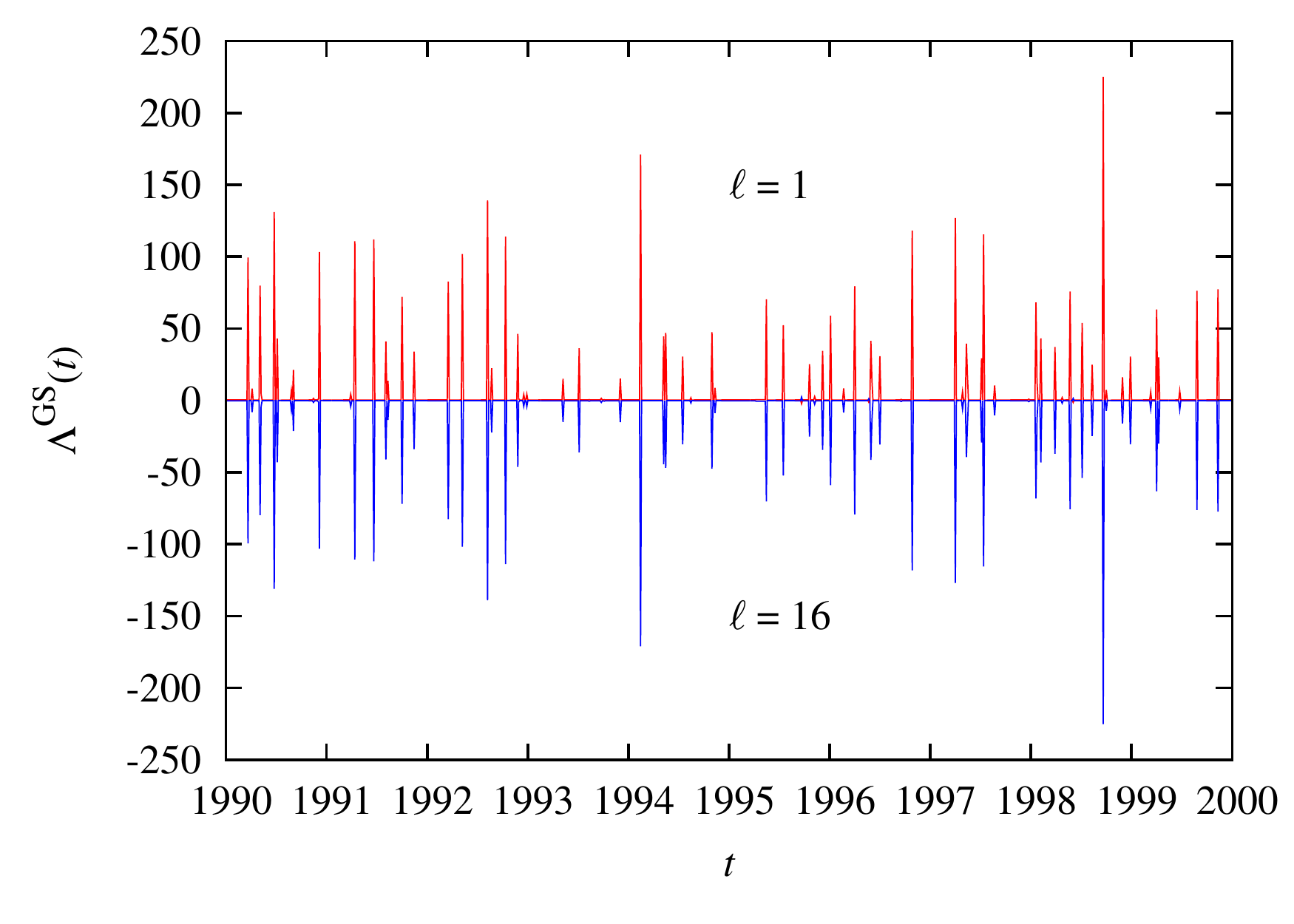}
\caption{(Color online) Test of the symplectic symmetry of Eq.~(\ref{gsfwd}) for a smooth hard-disk system
with $N = 4$. The phase space has 16 dimensions. The local GS exponents for $\ell = 1$ and $\ell = 16$ are 
plotted as a function of time $t$ along a short stretch of trajectory.
}
\label{symplectic}
\end{figure}

For later reference, we show in Fig.~\ref{symplectic} the time-dependence of the  local exponents for $\ell = 1$ and
$\ell = D = 4N = 16$ of a system consisting of four smooth hard disks.  The symplectic symmetry 
as given by  Eq.~(\ref{gsfwd}) is well obeyed. 

\section{Lyapunov modes}
\label{hydro}

Already the first simulations of the Lyapunov spectra for  hard disks revealed an interesting new step-like structure 
of the small positive and negative exponents in the center of the spectrum \cite{DPH_1996}, which is  due to
degenerate exponents.  A similar structure  was also found for hard dumbbell systems \cite{MPH_1998}, to which we come 
back in Sec.~\ref{dumbb}.  The explanation for this behavior lies in the fact that the perturbation vectors for these
exponents are characterized by coherent sinusoidal patterns spread out over the whole physical space (the simulation cell).
We have referred to these collective patterns as Lyapunov modes.
The modes are interpreted  as a consequence of a  spontaneous breaking of  continuous  symmetries and, hence,  
are  intimately connected with the zero modes  spanning the central manifold \cite{Szasz_Buch,Hthesis}. 
They appear as sinusoidal modulations of the zero modes in space -- with wave number $k \ne 0$ -- due to the spontaneous  
breaking of the continuous symmetries.  For $k \to 0$, the modes reduce to a linear superposition of the zero modes, and their
Lyapunov exponent vanishes. The experimentally observed wave vectors depend on the  size of the system and the 
nature of the boundaries (reflecting or periodic).

Our discovery of the Lyapunov modes triggered a number  of theoretical approaches.
Eckmann and Gat were the first to provide theoretical  arguments for the existence of the Lyapunov modes 
in transversal-invariant systems \cite{Eck}.  Their model did not have a dynamics in phase space but only an
evolution matrix in tangent space, which was  modeled as  a product of independent random
matrices.  In another approach, McNamara and Mareschal isolated the six hydrodynamic fields related to the invariants of the binary
particle collisions and the vanishing  exponents, and used a generalized Enskog theory to derive hydrodynamic evolution equations for
these fields.  Their solutions are the Lyapunov modes  \cite{McNamara}. In a more detailed
extension of this work restricted to small densities,  a generalized Boltzmann equation is used for the
derivation \cite{Mare}. de Wijn and van Beijeren pointed out  the analogy to the Goldstone mechanism of 
constructing (infinitesimal) excitations of the zero modes and the Goldstone modes of  fluctuating
hydrodynamics \cite{Goldstone,Forster}.   They used this analogy  to derive the modes
within the framework of kinetic theory \cite{Wijn_vB}.  With a few exceptions, a rather good agreement with the simulation results was achieved,  at least for low densities. Finally, Taniguchi, Dettmann, and Morriss approached the problem
from the point of view of periodic orbit theory \cite{TDM} and master equations \cite{TM_2002}. 

The modes were observed for hard-ball systems in one, two, and three dimensions 
\cite{DP_3d,Szasz_Buch,HPFDZ,TM_2003a,TM_2003b,Zabey}, for planar hard dumbbells \cite{Milano},  and also for one- and two-dimensional
soft particles \cite{Radons_Yang,Yang_Radons,FP_2005}.

A formal classification for smooth hard-disk systems has been given by Eckmann {\em et al.} \cite{Zabey}.  
If the $4N$ components of  a tangent vector are arranged according to
\begin{equation}
\delta {\bf \Gamma}\,=\,\left( \delta q_x,\delta q_y ; \delta p_x,\delta p_y  \right), \nonumber
\end{equation}
the six orthonormal zero modes, which  span the central manifold and are the 
generators of the  continuous symmetry transformations, are given by \cite{FHPH_2004,Zabey}
\begin{eqnarray}
\vec e_1 &=& \dfrac{1}{\sqrt{2 K}\,} \, ({p_x},{p_y};\,0,0) ,\nonumber \\
\vec e_2 &=& \dfrac{1}{\sqrt{N}\,}    \, (1,0              ;\ 0,0) ,\nonumber \\
\vec e_3 &=& \dfrac{1}{\sqrt{N}\,}    \, (0,1              ;\,0,0), \nonumber \\ 
\vec e_4 &=& \dfrac{1}{\sqrt{2 K}\,} \, (0,0               ;\,{p_x},{p_y} ) ,\nonumber \\
\vec e_5 &=& \dfrac{1}{\sqrt{N}\,} \, (0,0                  ;\,1,0) ,\nonumber \\
\vec e_6 &=& \dfrac{1}{\sqrt{N}\,} \, (0,0                  ;\,0,1) .\nonumber \\ \nonumber
  \end{eqnarray}
$\vec  e_1$ corresponds to a shift of the  time origin, 
$\vec e_4$ to a change of energy, $\vec e_2$ and $\vec e_3$  to a uniform translation 
in the $x$ and $y$ directions, respectively, and $\vec e_5$ and $\vec e_6$ to a
shift of the total momentum in the $x$ and $y$ directions, respectively.   
The six vanishing Lyapunov exponents associated with these modes have  indices $2N-2 \le i \le 2N+3$  in the center of the  spectrum.
Since, for small-enough $k$, the perturbation components of a particle obey
$\delta {\bf p} = C \delta {\bf q} $ for the unstable perturbations $(\lambda > 0)$, and 
$\delta {\bf q} = - C \delta {\bf p} $ for the stable perturbations $(\lambda < 0)$, where $C > 0$ is a number, one may restrict the classification 
to the $\delta {\bf q}$ part  of the modes and, hence, to the basis vectors $\vec e_1, \vec e_2, \vec e_3$ \cite{Zabey}.
Now the modes with a wave vector ${\bf k}$ may be classified as follows:
\begin{enumerate}
\item {\bf  Transverse modes}  (T)  are {\em divergence-free} vector fields consisting of  a superposition of sinusoidal 
modulations of $\vec e_2$ and $\vec e_3$.
\item {\bf Longitudinal modes}  (L)  are {\em irrotational} vector fields consisting of a superposition of sinusoidal modulations
of $\vec e_2$ and $\vec e_3$.
 \item {\bf Momentum modes}  (P) are vector fields consisting of sinusoidal modulations of  $\vec e_1$. Due to the
 random orientations of the particle velocities, a P mode is not easily recognized as fundamental mode in a simulation. 
 However, it may be numerically transformed into an easily recognizable periodic shape  \cite{Zabey,BP_2013}, as will
 be shown in Fig.~\ref{P10} below. 
\end{enumerate}
The subspaces spanned by these modes are denoted by $T ({\bf n})$,  $L ({\bf n})$, and  $P ({\bf n})$, respectively,
where the  wave vector for a periodic box with sides $L_x, L_y$  is  ${\bf k} = (2 \pi n_x / L_x, 2 \pi  n_y/ L_y)$, and
${\bf n} \equiv (n_x, n_y)$. $n_x$ and $n_y$ are integers.

For the same ${\bf k}$,  the values of the Lyapunov exponents for the
longitudinal and momentum modes coincide. These modes actually  belong to a combined
subspace $LP({\bf n}) \equiv L({\bf n}) \oplus P({\bf n})$. The dimensions of the subspaces  $T ({\bf n})$ and 
 $LP({\bf n})$ determine the multiplicity of the exponents associated with the T and LP modes. For a periodic
 boundary, say in $x$ direction, the multiplicity is 2 for the T, and 4 for the LP modes. For details we refer to
 Ref.~\cite{Zabey}.

The transverse modes are stationary in space and time, but the L and P modes are not  \cite{FHP_Congress,Zabey,CTM_2011}. 
 In the following, we only consider the  case of periodic boundaries.
Any tangent vector from an LP subspace observed in a simulation
is a combination of a pure L and a pure P mode. The dynamics of such an LP pair  may be represented as a rotation in a 
two-dimensional space with a constant angular frequency proportional to the wave number of the mode,
\begin{equation}
\omega_{\bf n} = v k_{\bf n}, \nonumber
\end{equation}
 where the L mode is continuously transformed  into  the P mode and vice versa.
 Since the P mode is modulated with the velocity field of the particles, the experimentally  observed mode pattern becomes periodically
 blurred when its character is predominantly P. Since during each half of a period, during which all 
 spacial sine and cosine functions change sign,
 the mode is offset in the direction of the wave vector ${\bf k}$, which gives it the appearance
 of a traveling wave with  an averaged  phase velocity $v$. If reflecting boundaries are used, standing waves are obtained.
The phase velocity $v$ is about one third of the sound velocity \cite{FHP_Congress}, but otherwise seems to be unrelated to it.

\begin{figure}[t]
\centering
\includegraphics[width=0.47\textwidth]{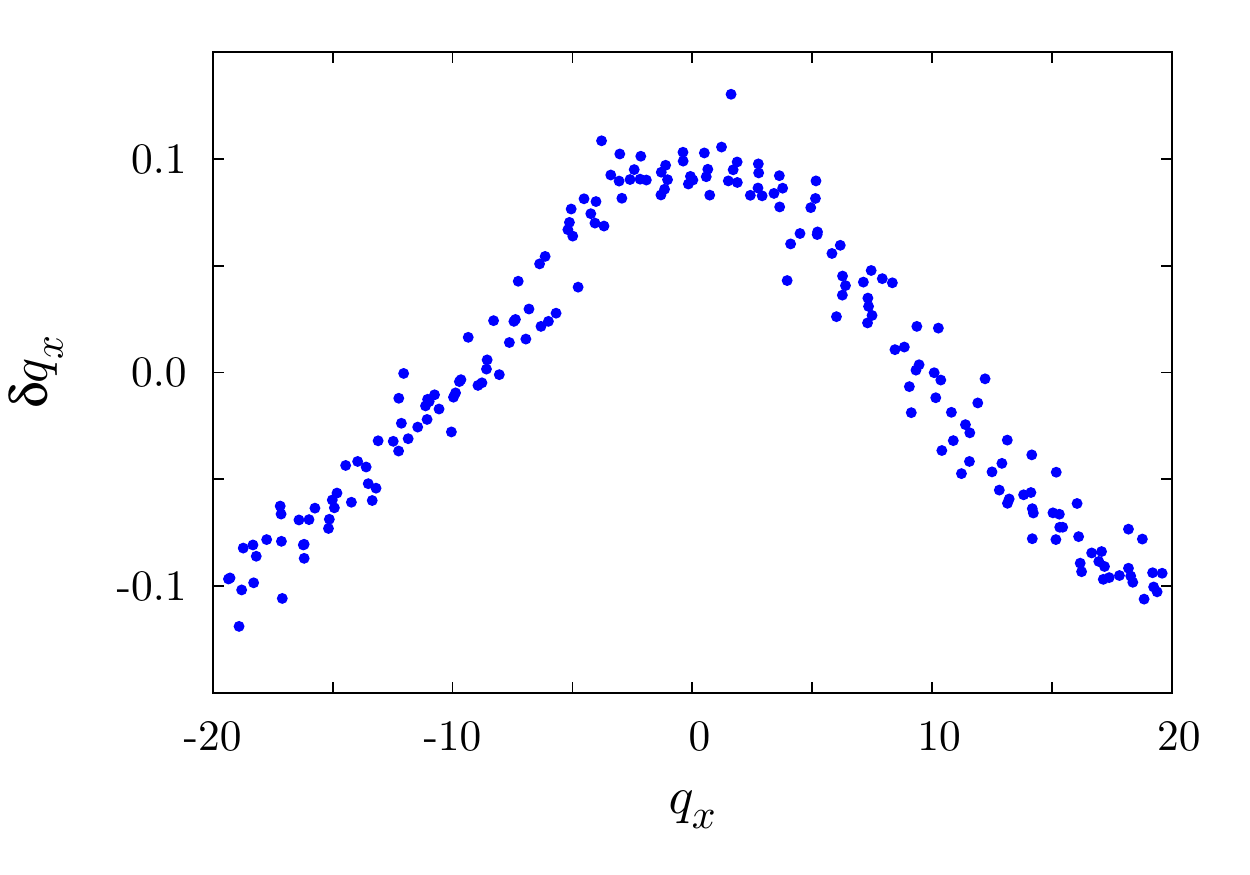} \\
\includegraphics[width=0.47\textwidth]{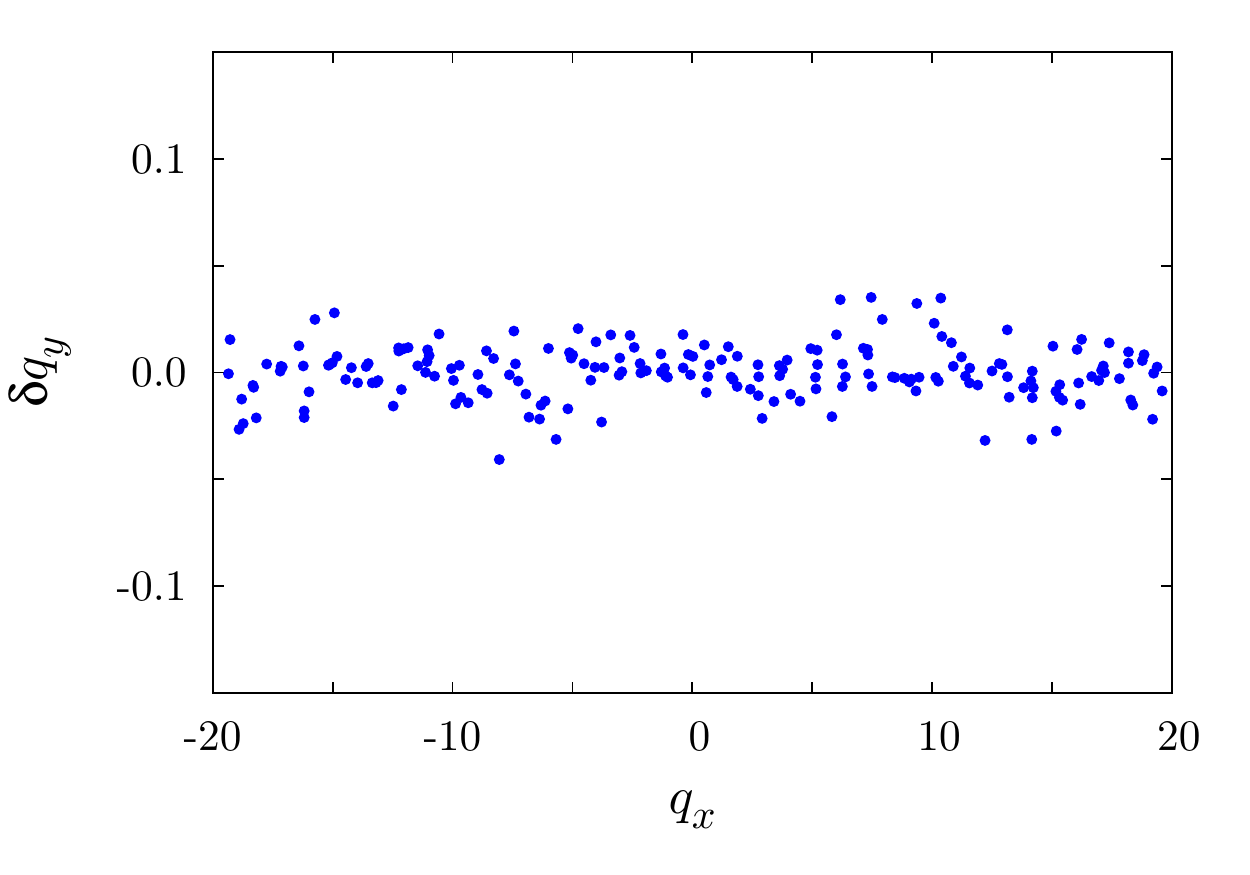} 
\caption{(Color online)  Reconstructed L(1,0) mode components proportional to  $\cos (2 \pi q_x /  L_x) $.
The system consists of 198 smooth hard disks in a periodic box. For details we refer to the main text.}
\label{L10}
\end{figure}

The basis vectors spanning the subspaces of the T and  LP modes may be reconstructed from the experimental data
with a least square method \cite{Zabey,BP_2010}. As an example, we consider the L(1,0) mode of a system
consisting of  198 hard disks at a density 0.7 in a rectangular periodic box with an aspect ratio 2/11. 
The LP(1,0) subspace includes the tangent vectors for $388 \le \ell \le 391$ with four identical Lyapunov exponents.
The mutually orthogonal basis vectors spanning the corresponding four-dimensional subspace are viewed as
vector fields in position space and   are given by \cite{Zabey}
\begin{equation}
L(1,0): \left(  \begin{array}{l} 1\\ 0 \end{array}  \right) \cos\left(\frac{2 \pi q_x}{ L_x }\right), \;\;
              \left(  \begin{array}{l} 1\\ 0 \end{array}  \right) \sin\left(\frac{2 \pi q_x}{ L_x }\right), 
\nonumber
\end{equation} 
\begin{equation}
P(1,0): \left(  \begin{array}{l} p_x\\ p_y \end{array}  \right) \cos\left(\frac{2 \pi q_x}{ L_x }\right), \;\;
              \left(  \begin{array}{l} p_x\\ p_y \end{array}  \right) \sin\left(\frac{2 \pi q_x}{ L_x }\right), 
\nonumber
\end{equation} 
where, for simplicity,  we have omitted the normalization. If the Gram-Schmidt vectors are used for the reconstruction, the components corresponding to the cosine  are  shown for L(1,0) in Fig.~\ref{L10}, for  P(1,0)  in Fig.~\ref{P10}. In the latter case, the mode 
structure is visible due to  the division by $p_x$ and $p_y$  as indicated.  This also explains the  larger scattering of the points.
The components proportional to the sine are analogous, but are not shown.  A related analysis may be carried out for the covariant vectors. 

Recently, Morriss and coworkers  \cite{CTM_2010,CTM_2011,MT_2013} considered in detail the tangent-space dynamics 
of the zero modes and of the mode-forming perturbations over a time $\tau$, during which many collisions occur.  In addition, they
enforced the mutual orthogonality of the subspaces of the modes and of their conjugate pairs (for which the Lyapunov  exponents only differ by  the sign), with the central manifold by invoking  an (inverse)  Gram-Schmidt procedure, which starts with the zero modes and works outward 
towards modes with larger exponents.  With this procedure  they were able to construct the modes in the limit of  large $\tau$ and  to obtain (approximate) values for the Lyapunov exponents for the first few Lyapunov modes (counted away from the null space) \cite{CTM_2010}. The agreement with simulation results is of the order of  a few percent for the  T modes,  and slightly worse for the LP modes. 

\begin{figure}[t]
\centering
\includegraphics[width=0.47\textwidth]{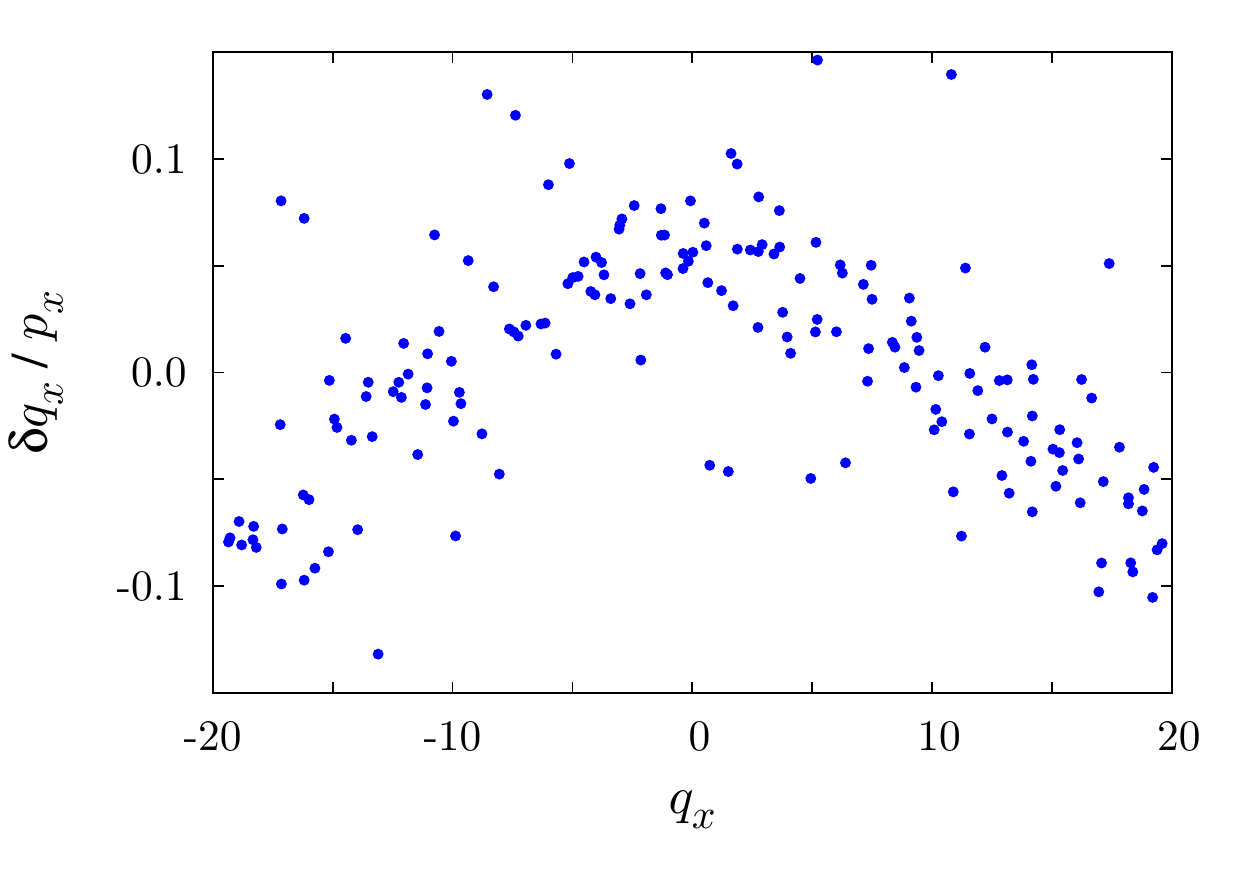}\\
\includegraphics[width=0.47\textwidth]{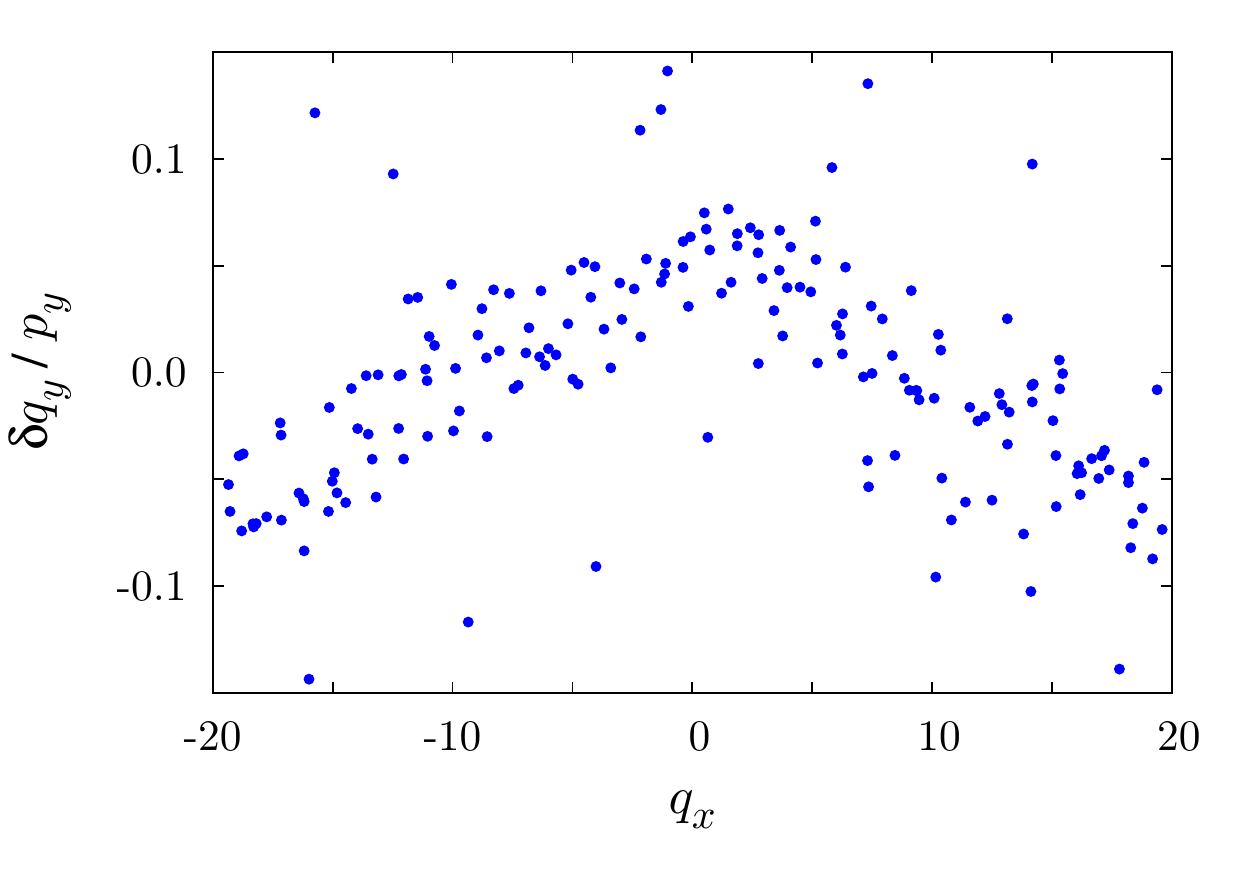}
\caption{(Color online)    Reconstructed P(1,0) mode  components proportional to $\cos (2 \pi q_x /  L_x) $
for the same LP subspace as in Fig.~\ref{L10}.}
\label{P10}
\end{figure}

 For the modes with a wave length  large compared to the inter-particle spacing, a  continuum limit for the perturbation vectors  leads to a set of partial differential equations for the perturbation fields, 
whose solutions give  analytical expressions for the modes in accordance with the boundary conditions applied. This procedure works for the stationary  transverse modes,  as well as for the time-dependent LP modes. For the latter, the partial differential equations for the continuous 
perturbation fields assume the form of a wave equation with traveling wave solutions \cite{CTM_2011}. 
For quasi-one-dimensional systems (with narrow boxes such that  particles may not pass each other),  the phase velocity is found to depend on the particle density via the mean particle separation  and the mean time between collisions (of a particle). For  fully two-dimensional models 
of hard disks,  the wave velocity becomes proportional to the collision frequency of a particle, and inversely proportional to the mean 
squared distance of a particle to all its neighbors in the first coordination shell   \cite{CTM_2011,MT_2013}. 
This is an interesting result, since it provides a long-looked-for connection between a mode property -- the wave velocity -- with other density-dependent microscopic quantities of a fluid.

\section{Rototranslation}
\label{roto}
\subsection{Rough hard disks}
\label{roughd}

Up to now we were concerned with simple fluids allowing only translational particle dynamics.  As a next step 
toward more realistic models, we consider fluids, for which the particles may also rotate and exchange energy 
between translational and rotational degrees of freedom. In three dimensions, the first model of that kind, rough hard spheres,
was already suggested by Bryan in 1894 \cite{bryan:1894}. Arguably, it  constitutes  the simplest model of a molecular 
fluid. It was later  treated by Pidduck \cite{pidduck:1922} and, in particular,  by 
Chapman and Cowling \cite{chapman:1953},
who derived the collision map in phase space for two colliding spheres with maximum possible roughness. 
The latter requires that the relative surface velocity at the point of contact of the collision partners is reversed,
leaving their combined (translational plus rotational) energy, and their combined linear and angular momenta invariant
\cite{Allen}.   The thermodynamics and 
the dynamical properties of that models were extensively studied by O'Dell and Berne \cite{Berne_I}. They also
considered generalizations to partial roughness in-between  smooth and maximally-rough spheres \cite{Berne_II,Berne_III,Berne_IV}. 
Here we consider the simplest two-dimensional version of that model, maximally-rough hard disks in a planar 
box with periodic boundaries. Explicit 
expressions for the collision maps in phase space and in tangent space may be obtained from our previous work 
\cite{vMP,BP_2013,Bdiss}. The coupling between translation and rotation is controlled  by a single dimensionless parameter  
$$\kappa = \frac{4I}{  m \sigma^2}, $$
where  $I$ is the principal moment of inertia for a rotation around an axis through the center,
and  $m$ and $\sigma$ denote the  mass and diameter of a disk, respectvely. $\kappa$ may take values between zero and one: $ 0,$ 
if all  the mass is concentrated in the center, $1/2$ for a uniform mass distribution, and $ 1$, 
if all the mass is located  on the perimeter of the disk. 

The rotation requires to add  angular momentum $J_i$ to the other independent variables ${\bf q}_i, {\bf p}_i$ of a particle $i$. 
The $J_i$ show up in 
the collision map \cite{chapman:1953,vMP,BP_2013}, and their perturbations $\delta J_i$  affect the tangent space dynamics
\cite{vMP,BP_2013}. If one is only interested in the global Lyapunov exponents, this is all what is needed, since it  constitutes  an autonomous system. 
We refer to it as the $J$-version of the rough-disk model. Its phase-space dimension $D = 5N$. Since $D$ may be an odd number, it is obvious that one cannot ask questions about the symplectic nature of the model. For this it is necessary to include  
also the disk orientations $\Theta_i$ conjugate to the angular momenta in the list of
independent variables. The angles do not show up in the collision map of the reference trajectory, but their perturbations,
$\delta \Theta_i$, affect the collision map in tangent space.  Now the phase space has 6N dimensions. We refer to this representation as 
the $J \Theta$-version.

Before entering the discussion of the Lyapunov spectra for large systems,  let us  clarify the number of vanishing exponents first. 
\begin{figure}[t]
\centering
\includegraphics[width=0.40\textwidth]{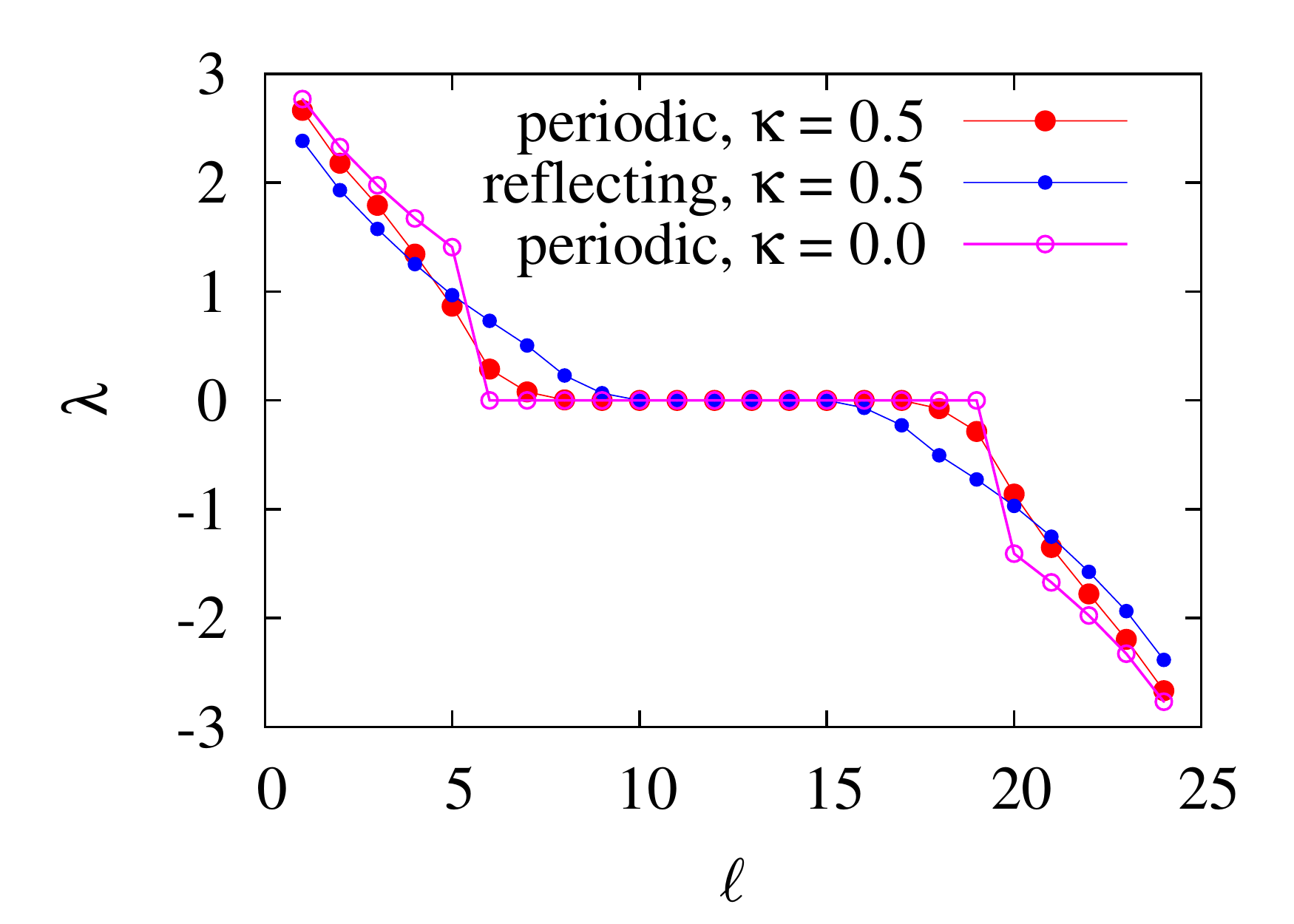}
\caption{(Color online) Full Lyapunov spectra in $J \Theta$-representation for four rough disks $(\kappa = 0.5)$ respective 
smooth disks $(\kappa = 0)$ with boundary conditions as indicated.  It is interesting to note that the equipartition 
theorem does not hold for such small systems.
For  $\kappa = 0.5$, the translational and rotational kinetic energies add according to $3.59 + 2.41 = 6.$
For the simulation with $\kappa = 0$, a total translational kinetic energy equal to two was used. 
}
\label{zero_exponents}
\end{figure}
 As an illuminating case, we plot in Fig.~\ref{zero_exponents} some  spectra for the  $J \Theta$-version of a 
 rough disk system with only 4 particles and 24 exponents. For positive $\kappa$, when translation and rotation
 are intimately coupled, the number of vanishing exponents is determined by the intrinsic symmetries  of
 space and time-translation invariance and by the number of particles. For {\em periodic boundaries} in $x$ and $y$ directions
 (full red dots in   Fig.~\ref{zero_exponents}), space-translation invariance contributes four, time-translation 
 invariance another two vanishing exponents. Furthermore, the invariance of the collision map with respect to an arbitrary rotation 
 of a particle adds  another $N$ zero exponents, four in our example. Altogether,  this gives 10 vanishing
 exponents, as is demonstrated in Fig.~\ref{zero_exponents} by the red points. For {\em reflecting} boundaries in $x$ and $y$ 
 directions (small blue dots in  Fig.~\ref{zero_exponents}) there is no space-translation invariance, and the total number 
 of vanishing exponents is limited to six. If $\kappa$ vanishes, the rotational energy also vanishes, and all  variables connected with 
 rotation cease to be chaotic  and contribute another $2N$ zeroes to the spectrum. Together with the 
 intrinsic-symmetry contributions, this amounts to 
 $2N+6 = 14$ vanishing exponents for periodic boundaries (open purple circles in  Fig.~\ref{zero_exponents}),
 respective $2N + 2 = 10$ for reflecting boundaries (not shown).  Note that we have plotted the full spectra including the
 positive and negative branches in this figure. For the $J$-version, and for an odd number of particles,  the discussion 
 is slightly more involved for which we refer to Ref.~\cite{BP_2013}.

Next we turn to the discussion of large systems, for which deviations from equipartition are negligible.  We use reduced units for which
$m$,  $\sigma$ and the kinetic translational (and rotational) temperatures  are unity. Lyapunov exponents are given in units of $\sqrt{K/Nm\sigma^2}.$ The system we first  consider in the following consists of 
$N = 88$ rough hard disks with a density $\rho = 0.7$ in a periodic box with an aspect ratio $A = 2/11$ \cite{BP_2013}.
For comparison, spectra for 400-particle systems are given in Ref.~\cite{vMP}. Since we are here primarily  interested in the global exponents,
the simulations are carried out with the $J$-version of the model. 

\begin{figure}[t]
\centering
\includegraphics[width=0.48\textwidth]{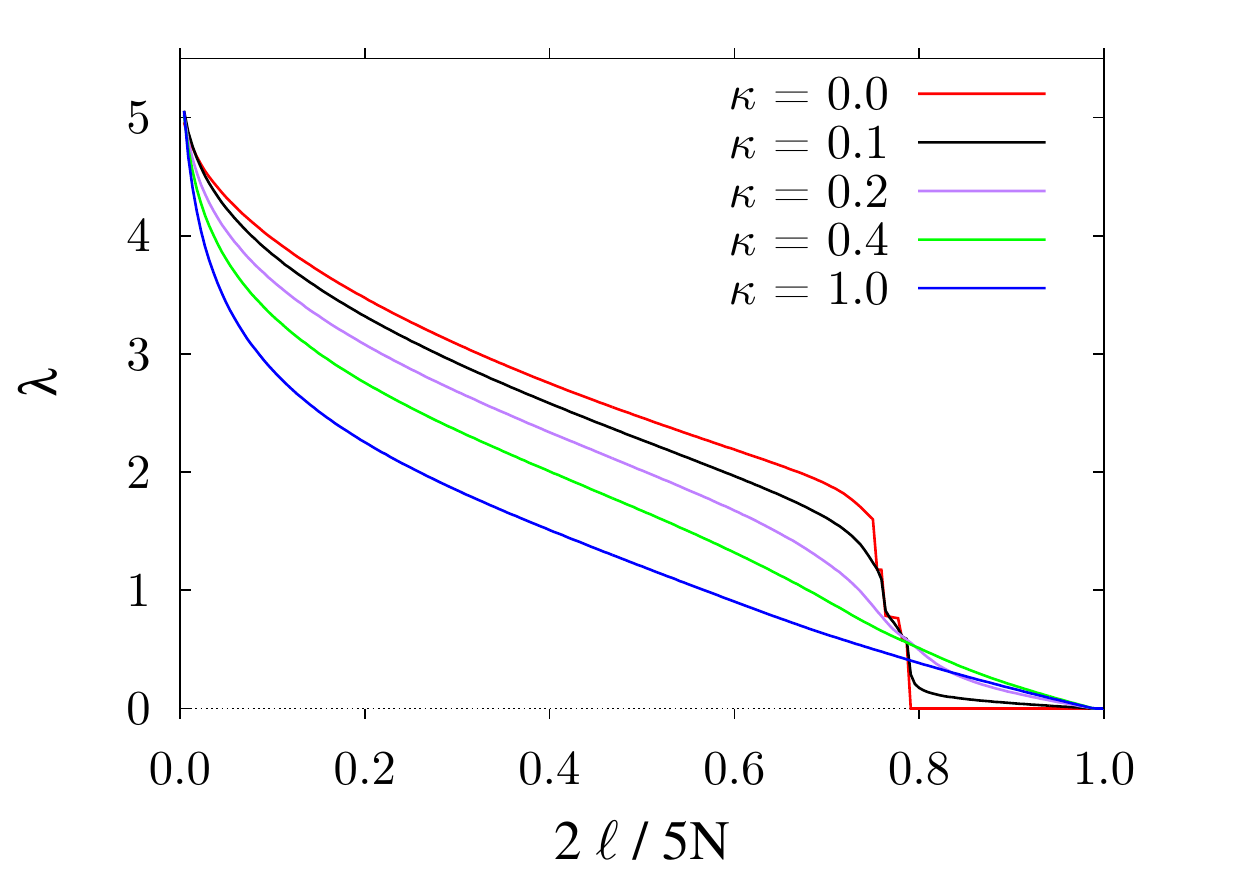}
\caption{(Color online) Lyapunov spectra (positive branches only) for a system of 88 rough hard disks
for various coupling parameters $\kappa$ as indicated by the labels \cite{BP_2013}. 
The density $\rho = 0.7$. The periodic box has an aspect ratio of 2/11.
}
\label{spectrum_rough}
\end{figure}
\begin{figure}[t]
\centering
\includegraphics[width=0.48\textwidth]{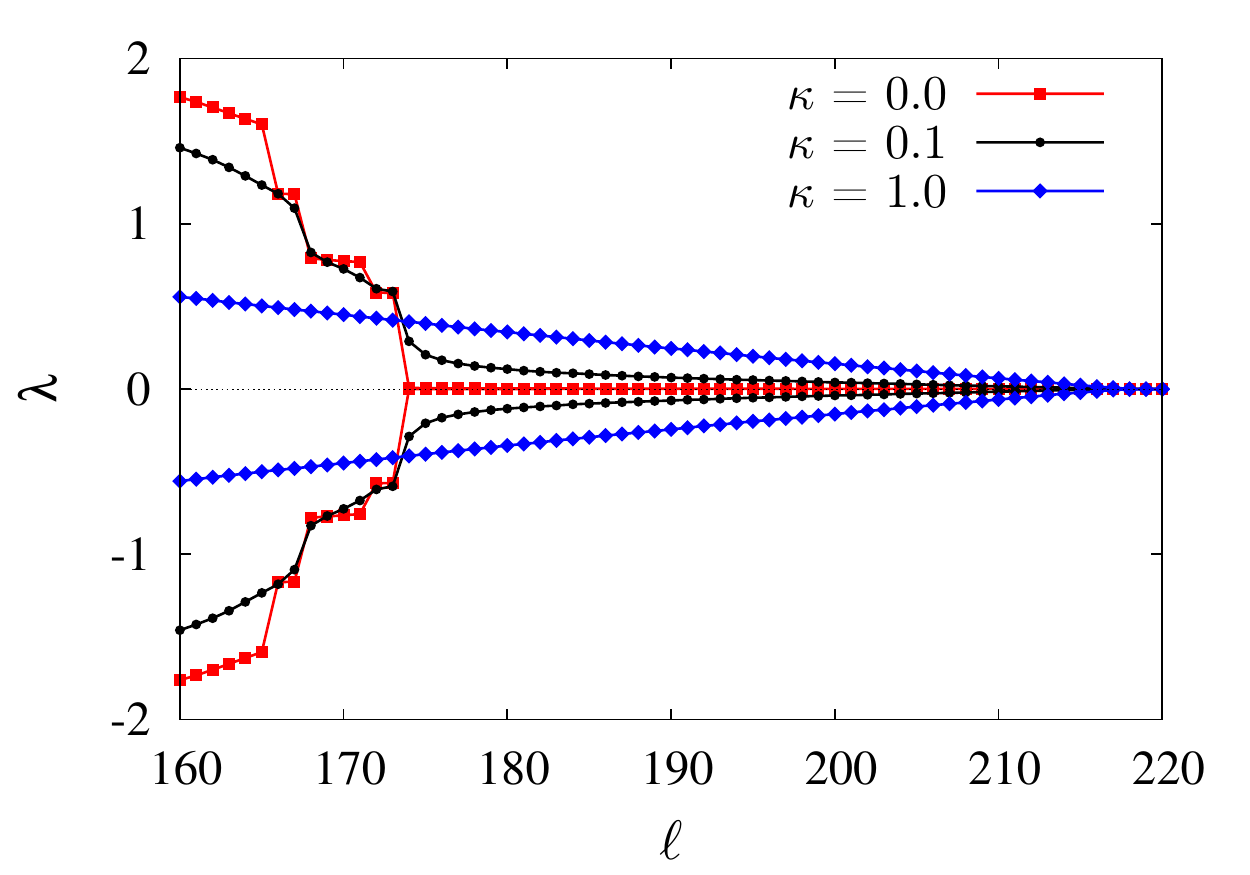}
\caption{(Color online) Enlargement of the regime of small Lyapunov exponents for some spectra of
Fig.~\ref{spectrum_rough}. Also negative exponents are displayed such that conjugate pairs, $\lambda_{\ell}$ and
$\lambda_{5N+1-\ell}$, share the same index $\ell$ on the abscissa. 
}
\label{center_rough}
\end{figure}
In Fig.~\ref{spectrum_rough} we show the positive branches of the Lyapunov spectra for a few selected values of $\kappa$ \cite{BP_2013}. Although the exponents are only defined  for integer $\ell$, smooth lines are drawn for clarity, and a reduced index
is used on the abscissa. There are $5N = 440$ exponents, of which the first half constitute the positive branch 
shown in the figure. An enlargement of the small exponents near the center of the spectrum is
provided in Fig.~\ref{center_rough} for selected values of $\kappa$. There we have  also included the negative branches in such a way to emphasize conjugate pairing.  
For $\kappa = 0$, which  corresponds to freely rotating smooth disks without roughness, the Lyapunov spectrum is identical 
to that of pure smooth disks without any rotation, if $N = 88$ vanishing exponents  are added to the spectrum. 
The simulation box is long enough (due to  the small aspect ratio)
for Lyapunov modes to develop along the  $x$ direction. This may be verified by the
step structure of the spectrum due to  the small degenerate exponents. If $\kappa$ is increased, the Lyapunov modes quickly disappear,
and hardly any trace of them is left  for $\kappa = 0.1$. For small positive $\kappa$ the spectrum is decomposed into a rotation-dominated 
regime for $2N < \ell < 3N$  and a translation-dominated regime for $ \ell \le 2N$ and $\ell \ge 3N$.
It is, perhaps,  surprising that, with the exception of the maximum exponent $\lambda_1$,  already a small admixture 
of rotational motion significantly reduces the translation-dominated exponents, whereas the rotation-dominated exponents 
only gradually increase.  

\begin{figure}[b]
\centering
\includegraphics[width=0.46\textwidth]{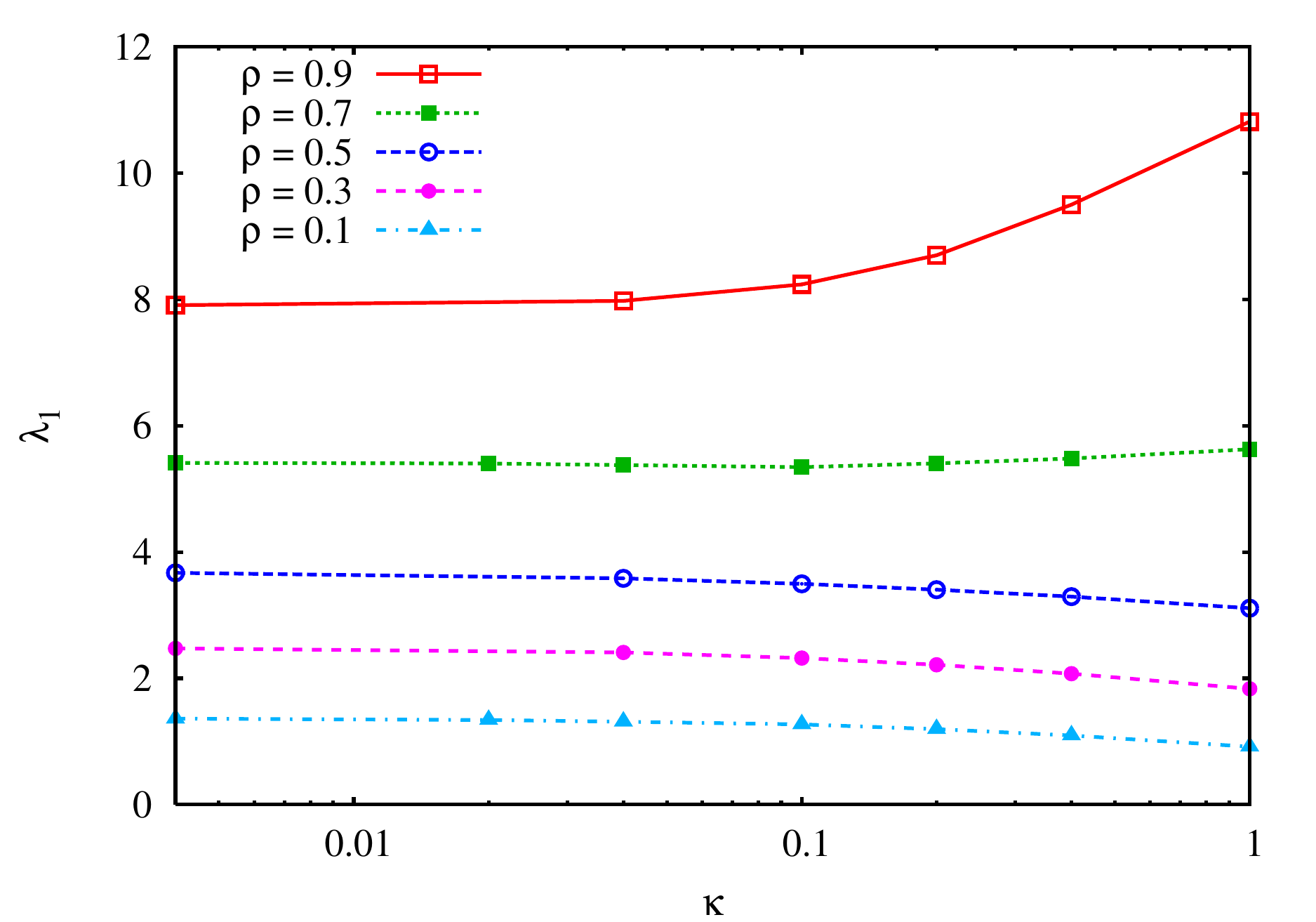}
\caption{(Color online) Dependence of the maximum Lyapunov exponent on the coupling parameter $\kappa$
for the rough-disk fluid at various densities. From Ref.~\cite{vMP}. 
}
\label{max_kappa}
\end{figure}

\begin{figure}[t]
\centering
\includegraphics[width=0.46\textwidth]{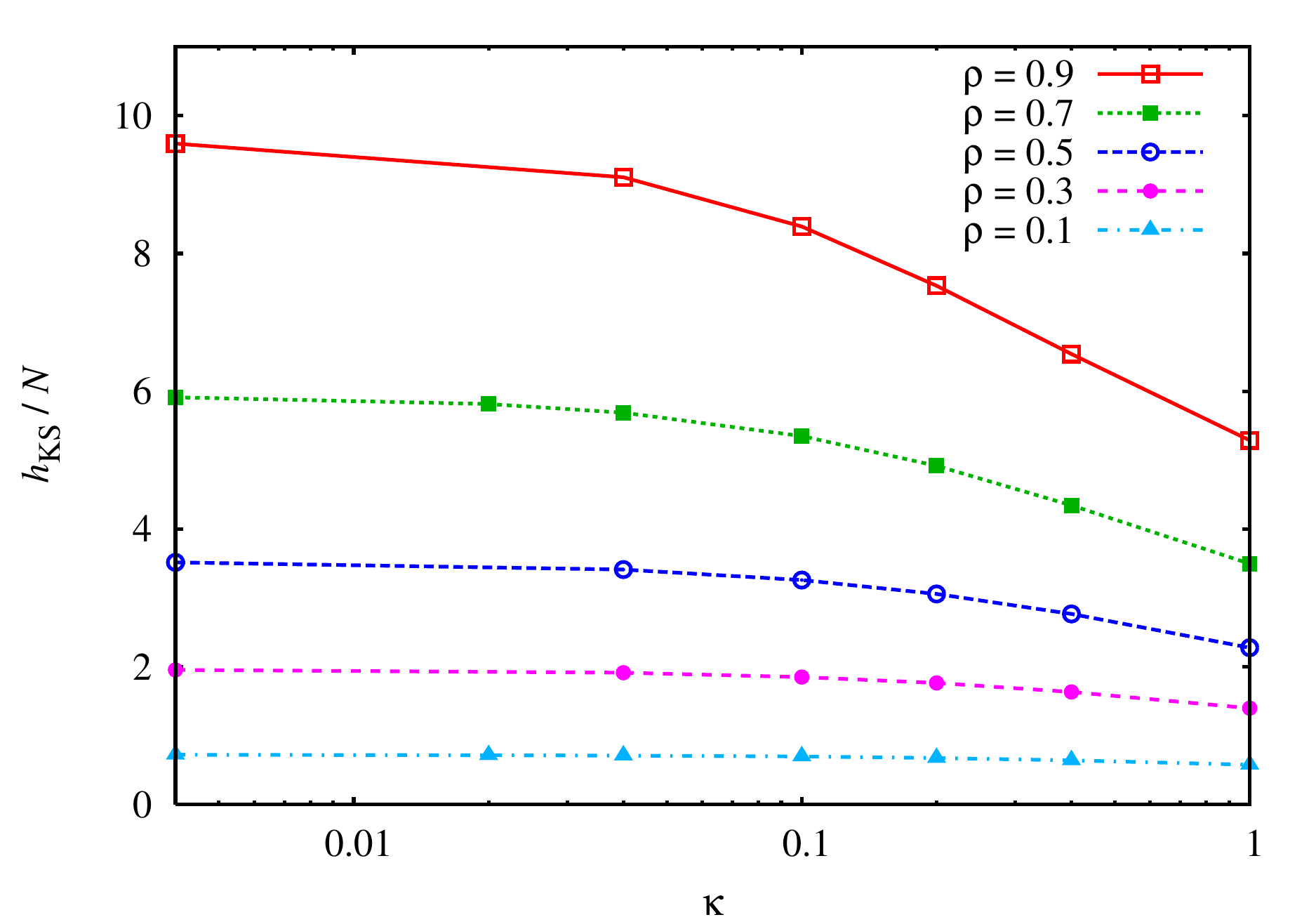}
\caption{(Color online) Dependence of the Kolmogorov-Sinai entropy per particle on the coupling parameter $\kappa$
for the rough-disk fluid at various densities. From Ref.~\cite{vMP}
}
\label{KS_kappa}
\end{figure}

To study this further, we plot in the Figs.~\ref{max_kappa} and~\ref{KS_kappa} the isothermal $\kappa$-dependence  of the 
maximum exponent, $\lambda_1$, and of the 
Kolmogorov-Sinai entropy per particle, $h_{KS}/N$,  respectively,   for various fluid densities as indicated by the labels. 
These data are  for a system consisting of 400 rough disks  \cite{vMP}. 
$\lambda_1$  and, hence, dynamical chaos tends to decrease with $\kappa$ for low densities, and  always increases for large densities.  
At the same time, the KS-entropy always decreases with $\kappa$. This means that mixing becomes less effective and the
time for phase-space mixing goes up,  the more the rotational degrees of freedom interfere with the translational dynamics.

\begin{figure}[t]
\centering
\includegraphics[width=0.47\textwidth]{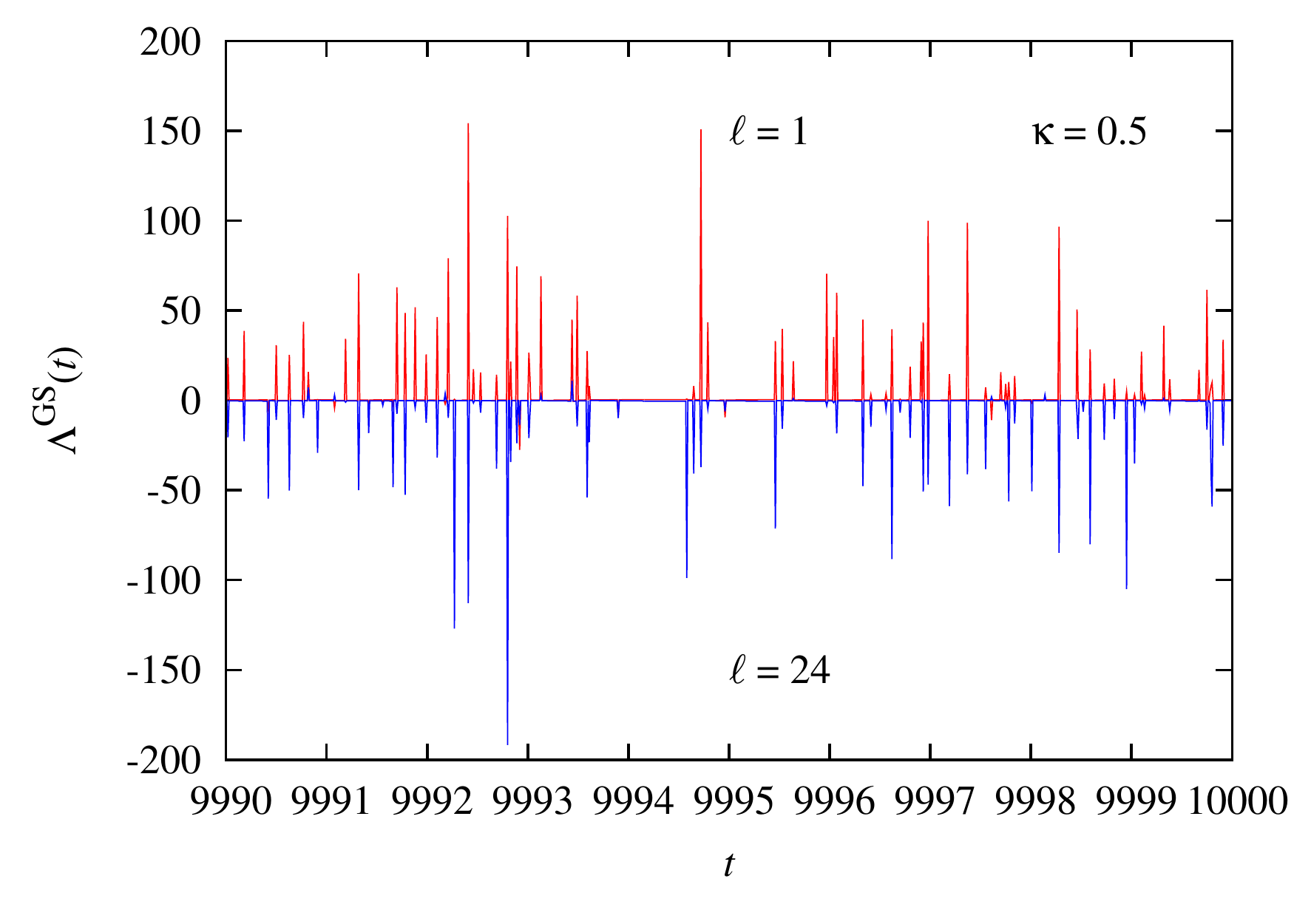}
\caption{(Color online) Demonstration of the lack of symplectic symmetry (as formulated in  Eq.~(\ref{gsfwd})) for a rough  hard-disk system with $N = 4$ particles. The simulation is carried out with the $J \Theta$-representation of the model, which includes the disk orientations, and for which  the phase space has 24 dimensions. The coupling parameter $\kappa = 0.5$.  The local GS-exponents for
$\ell = 1$ and $\ell = 24$ are plotted as a function of time $t$ for a short stretch of trajectory.}
\label{rough_symplectic}
\end{figure}

 Before leaving the rough hard-disk case, a slightly disturbing  observation for that system should be mentioned.  In Fig.~\ref{symplectic}
 we have demonstrated that the local GS Lyapunov exponents of the smooth hard-disk model display symplectic
 symmetry. Unexpectedly,  for the rough hard-disks this is not the case.  This is shown in Fig.~\ref{rough_symplectic}.
where the maximum and minimum local exponents do not show the expected symmetry, and neither do the other
conjugate pairs which are not included in the figure.  If the collision map for two colliding particles in tangent space
is written in matrix form, $\delta {\bf \Gamma}' = {\cal M} \delta {\bf \Gamma},$ where $\delta {\bf \Gamma}$ 
and   $\delta {\bf \Gamma}' $  are the perturbation vectors  immediately before and after a collision, respectively,  
the matrix ${\cal M}$ does not obey the symplectic condition ${\cal S}^{\dagger} {\cal M} {\cal S } = {\cal M}.$ Here,
${\cal S}$ is the anti-symmetric symplectic  matrix, and  $^{\dagger}$ means transposition.  For a more extensive
discussion we refer to Refs.~\cite{BP_2013,Bdiss}.

\subsection{Hard dumbbells}
\label{dumbb}

\begin{figure}[b]
\centering
\includegraphics[width=0.35\textwidth]{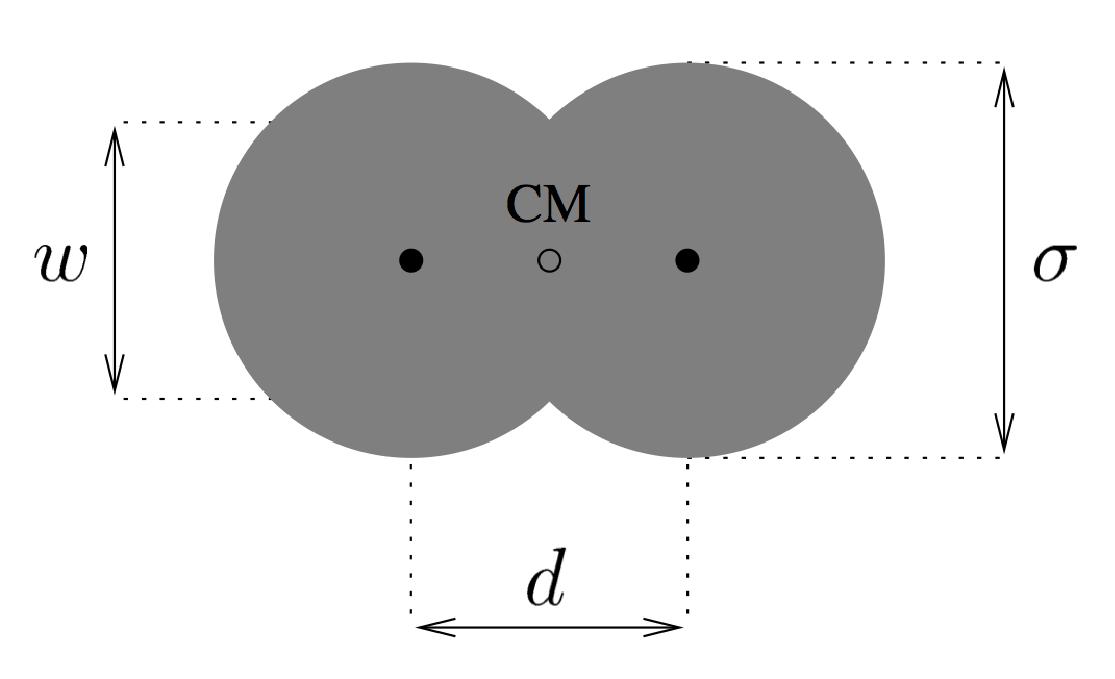}
\caption{ Geometry of a hard dumbbell diatomic. All quantities for this model are given in terms of
reduced quantities, for which the disk diameter $\sigma$, the molecular mass $m$, and the total kinetic energy
per molecule, $K/N,$ are equal to unity. The molecular anisotropy is defined by $d/\sigma$.}
\label{dumbbell_geometry}
\end{figure}

A slightly more realistic model for linear molecules are  ``smooth'' hard dumbbells, whose geometry is shown in
Fig.~\ref{dumbbell_geometry}.  As with hard disks, the dynamics is characterized by hard encounters with impulsive forces perpendicular to the surface. Between collisions, the particles translate and rotate  freely, which
makes the simulation rather efficient \cite{AI_1987,TS_1980,Bellemans_1980}. In the following we restrict to the 
planar version of this model, for which the ''molecule'' consists of two fused disks as shown in Fig.~\ref{dumbbell_geometry}.
The state space is spanned by the center-of-mass (CM) coordinates ${\bf q}_i$, the 
translational  momenta  ${\bf p}_i$, the orientation angles $\Theta_i$, and the angular momenta $J_i$,   $i = 1,2,\dots,N$,
and has $6N$ dimensions. 
The Lyapunov spectra for this model were first computed by Milanovi{\'c} {\em et al.} 
\cite{MPH_1998,MPH_chaos,Milano,Mdiss}.  Related work for  rigid diatomic molecules  interacting with
soft repulsive site-site  potentials was carried out by Borzs\'ak {\em et al.} \cite{Borzsak} and 
Kum {\em et al.} \cite{Kum}.

Here we restrict the discussion to homogeneous dumbbells, for which the molecular mass $m$ is uniformly distributed
over the union of the two disks.  The moment of inertia for rotation around the center of mass  becomes
\begin{equation}
     I(d \le \sigma) =  
            \frac{m \sigma^2}{4} \frac{ 3 d w + [d^2 + (\sigma^2/2)]
           [\pi + 2 \arctan(d/w)]}{ 2 d w + \sigma^2 
           [\pi + 2 \arctan(d/w)]},
\nonumber
\end{equation}
and
\begin{equation}
I(d > \sigma) = \frac{m}{4}\left[\frac{\sigma^2}{2}+d^2\right], \nonumber
\end{equation}
 where $w = \sqrt{\sigma^2 - d^2}$ is the molecular waist.
 $I(d)$ monotonously increases with $d$.  For $d\to 0$ it  converges to that
of a single disk with mass $m$ and diameter $\sigma$, and with a homogeneous
mass density.
Results for other mass distributions may be found  in Ref.~\cite{Milano}.
$d$ plays
a similar role as the coupling parameter $\kappa$ for the rough-disk model.

\begin{figure}[t]
\centering
\includegraphics[width=0.45\textwidth]{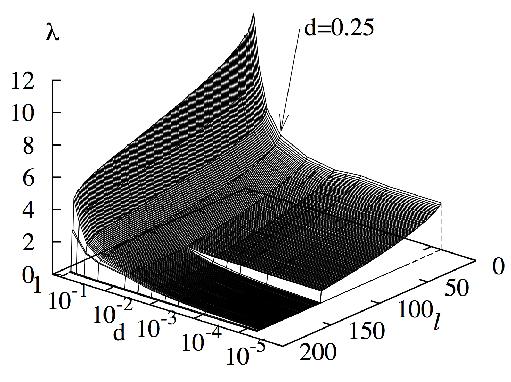}
\caption{  Anisotropy dependence of the Lyapunov spectra for a system of 64 planar hard dumbbells 
in a square box with periodic boundaries. $d$ is the molecular anisotropy (in reduced units).  Only the positive branches of the 
spectra with $3N = 192 $ exponents are shown. The number density is 0.5.  The Lyapunov index is denoted by $l$. 
(From Ref.~\cite{MPH_1998}). 
 } 
\label{surface}
\end{figure}

The results are summarized in  Fig.~\ref{surface}, where  the Lyapunov spectra, positive branches only,  
for a system of 64 dumbbells in a periodic square box at an intermediate gas density of $0.5$ are shown for various 
molecular anisotropies $d$. The system is too small for Lyapunov modes to be clearly visible
(although a single step due to a transverse mode may be observed for $d > 0.01$ even in this case).   
Most conspicuous, however,  is the widening gap in the spectra between the indices $l = 2N-3 = 125$ and $l = 126$
for $ d < d_c \approx 0.063.$ It separates the translation-dominated exponents ($1 \le l \le 125$) from the rotation-dominated exponents
($126 \le l \le 189$). The gap disappears for $d > d_c$.

\begin{figure}[ht]
\centering
\includegraphics[width=0.48\textwidth]{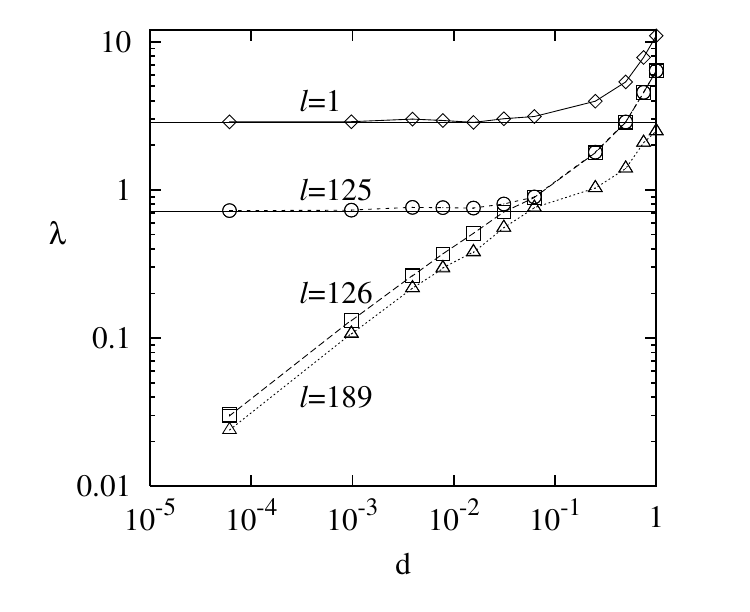}
\caption{ Anisotropy dependence of  selected Lyapunov exponents, labelled by their indices $l$,
  for the 64-dumbbell system summarized in Fig.~\ref{surface}.  The horizontal lines indicate the values for corresponding exponents of  a smooth hard-disk gas.
From Ref.~\cite{MPH_1998}.
} 
\label{aniso}
\end{figure}

It was observed in Ref.~\cite{MPH_1998} that for $d < d_c$ the perturbation
vectors associated with the translation- and rotation-dominated exponents  predominantly  point  into the subspace
belonging to center-of-mass translation and molecular rotation, respectively, and very rarely  rotate into a direction belonging to the other subspace. For anisotropies $d > d_c$, however, one finds
that the offset vectors for {\em all} exponents spend comparable times in both subspaces, which is taken as an indication of strong
translation - rotation coupling. $d_c$ is found to increase with the density. The anisotropy dependence
of some selected Lyapunov exponents is
shown in Fig.~\ref{aniso}. The horizontal lines for $l=1$ and $l=2N-3=125$ indicate the values
of the maximum and smallest positive exponents  for a system of 64 hard disks with the same density,
 to which the respective exponents of the dumbbell system converge with $d \to 0$. The smooth hard-disk data
 were taken from Ref.~\cite{DPH_1996}.

It was shown in the Refs.~\cite{MPH_1998,Milano} that two ``phase transitions'' exist for the hard dumbbells with an anisotropy of $d=0.25$,
a first at a number density $n_1 = 0.75$ from a  fluid to a rotationally-disordered solid, and a second, at $n_2 \approx 0.775$, from the 
orientationally-disordered solid to a crystal with long-range orientational order. Both transitions were observed by computing
orientational correlation functions. The first transition
at $n_1$ makes itself felt by large fluctuations of the Lyapunov exponents during the simulation and a slow convergence 
\cite{Mdiss}, but does not show up in the density dependence of the Lyapunov exponents. The second transition at $n_2$ does give 
rise to steps in the density-dependence curves of both $\lambda_1$ and  of $\lambda_{189}$  (which is the smallest positive exponent). 
These steps are even more noticeable when viewed as a function of the collision frequency instead of the density,
and the step is  significantly larger for  $\lambda_{189}$ than for $\lambda_1$. This indicates that the collective long-wavelength perturbations are much stronger affected by the locking of dumbbell orientations than the large exponents, which -- due to localization -- 
measure only localized dynamical events.  Here, dynamical systems theory offers a new road to an understanding of collective phenomena 
and phase transitions, which has not been fully exploited yet. 

\begin{figure}[t]
\centering
{\vspace{-8mm}\includegraphics[width=0.4\textwidth]{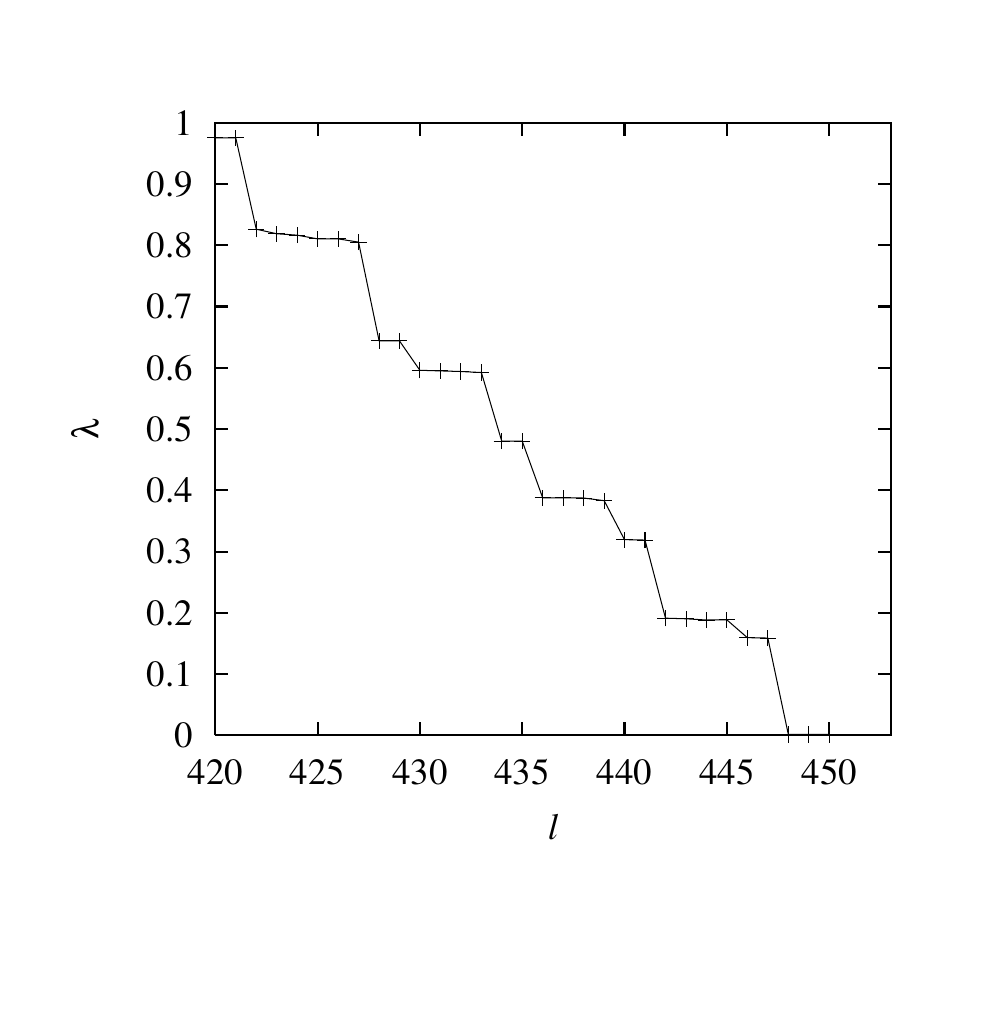}}
{\vspace{-15mm}
\caption{Magnification of the lower
	part of the Lyapunov spectrum for a system consisting of $N=150$
	homogeneous dumbbells with a molecular anisotropy
	$d=0.25$. The density $n=0.5$, and the aspect ratio of the box is $1/32$. The
	Lyapunov exponents are given in units of
	$\sqrt{(K/N)/(m\sigma^2)}$, where $K$ is the total (translational and rotational) kinetic
	energy, $m$ is the dumbbell mass, and $\sigma$ is the diameter
	of the two rigidly connected disks forming a dumbbell. The complete spectrum contains 
	900 exponents.}
\label{low}}
\end{figure}

Unlike the rough hard disks, the hard dumbbell systems generate well-developed Lyapunov modes for their small exponents.
Instructive examples  for $N = 400$ homogeneous hard dumbbells 
with an anisotropy $d = 0.25$ are given in the Refs.~\cite{Milano,Mdiss}. The classification of the modes 
in terms of transversal (T) and longitudinal-momentum (LP) modes is completely 
analogous to that for hard disks \cite{Zabey}, although some modifications   due to the presence of rotational degrees of freedom 
have not yet been worked out in all detail.  In Fig.~\ref{low}, we show, as an example, the smallest positive exponents of a
system of $N = 150$ dumbbells with a density $n = 0.5$ in an elongated (along the $x$ axis)  periodic box with  an aspect ratio $A = 1/32$
\cite{Mdiss}.
\begin{figure}[b]
\centering
\includegraphics[width=0.44\textwidth]{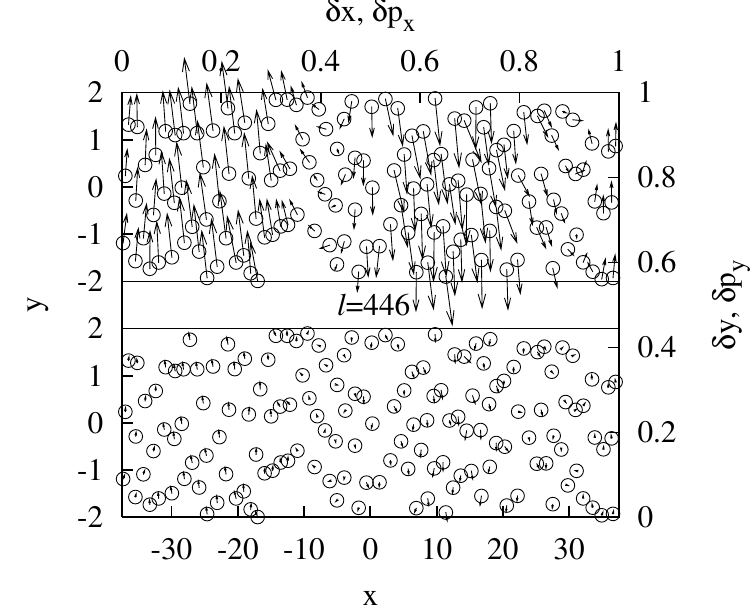}
\caption{Transverse mode pattern for the perturbation vector $\delta {\bf \Gamma}_{446}$
for an instantaneous configuration of the 150-dumbbells system with the Lyapunov
spectrum shown in Fig.~\ref{low}.  The circles denote the center-of-mass positions of the dumbbells, and the arrows indicate the position perturbations  $(\delta x, \delta y)$ (top) and momentum perturbations $(\delta p_{x}, \delta p_y)$ (bottom) of the particles. The system is strongly contracted along the x-axis.}
\label{trans_446}
\end{figure}
Only modes with wave vectors parallel to $x$ develop, which facilitates the analysis. The spectrum displays  the 
typical step structure due to the mode degeneracies  with multiplicity two, for the T modes, and four, for the LP modes.      
For this system, typical instantaneous mode patterns are shown in the
Figs.~\ref{trans_446} and~\ref{trans_443}, the former for a T mode with 
$l = 446$, the latter for an LP mode with $l = 443$.  In both cases, the wavelength is equal to the box length in $x$ direction. 
Please note that the system appears strongly contracted along the $x$ axis. 

\begin{figure}[t]
\centering
\includegraphics[width=0.44\textwidth]{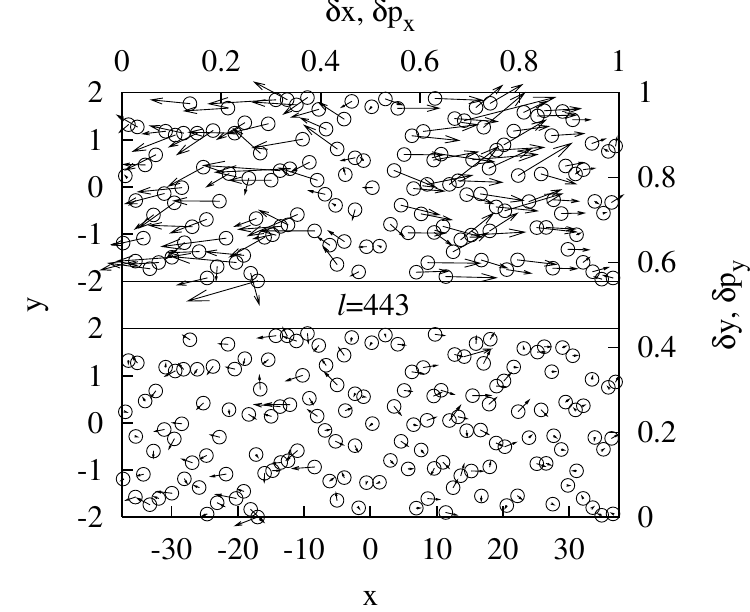}
\caption{As in the previous Fig.~\ref{trans_446}, but for the longitudinal-momentum 
mode belonging to the  perturbation vector $\delta {\bf \Gamma}_{443}.$
}\label{trans_443}
\end{figure}

\section{Stationary particle systems out of equilibrium}
\label{nonequilibrium}
One of the most interesting -- and  historically  first -- studies of Lyapunov spectra for many-particle systems 
is for stationary  non-equilibrium states \cite{HHP_1987,PH88,PH_1989}. If a system 
is driven away from equilibrium by an
external or thermal force, the irreversibly generated heat needs to be removed and transferred
to a heat bath. Otherwise the system would heat up and would never reach a steady state.
Such a scheme may assume the following equations of motion,
\begin{eqnarray}
\dot{{\bf q}}_i &=& {\bf p_i}/m_i +  {\cal{Q}}(\{{\bf q}\},\{{\bf p}\}) X(t) \nonumber\\
\dot{{\bf p}}_i &=& - \partial \Phi/\partial {\bf q}_i + {\cal{P}}(\{{\bf q}\},\{{\bf p}\}) X(t)  - s_i\zeta {\bf p}_i,  \label{thermo}\\ 
\nonumber
\end{eqnarray}
where ${\cal{Q}} X$ and ${\cal{P}} X$ represent the driving, and $-\zeta {\bf p}_i$ the thermostat or heat bath
acting on a particle $i$ selected with the switch $s_i \in \{ 0,1\}$.
$\zeta$ fluctuates  and may be positive or negative,  whether 
kinetic energy has to be extracted from -- or added to -- the system. Averaged over a long trajectory, 
$\left\langle \zeta \right\rangle$ needs to be positive.
$\zeta(\{{\bf q}\},\{{\bf p}\})$ may be a
Lagrange multiplier, which minimizes the thermostatting force according to Gauss' principle of least constraint
and keeps either the kinetic energy or the total internal energy a constant of the motion. Or it may be a single
independent variable in an extended phase space such as for the Nos\'e-Hoover thermostat.
There are excellent monographs  describing these schemes \cite{EM_2008,HH_2012}.
They are  rather flexible and allow for any number of  heat baths.  Although it is not possible to
construct such thermostats in the laboratory, they allow very efficient non equilibrium molecular dynamics (NEMD)
simulations and are believed to provide an accurate description of the nonlinear transport  involved \cite{Ruelle:1999}. 

Simple  considerations lead to the following thermodynamic relations for the stationary non-equilibrium state generated in this way after 
initial transients have decayed:
\begin{eqnarray}
&&\left\langle \frac{ d \ln \delta V^{(D)} }{dt }  \right\rangle  = \left\langle  \frac{\partial}{\partial {\bf \Gamma}} \cdot 
              \dot{{\bf \Gamma}}  \right\rangle \nonumber  = \sum_{\ell =1}^{D} \lambda_{\ell}
          =   - d \sum_{i=1}^N s_i \left\langle \zeta \right\rangle \nonumber \\   
       &=& -\frac{1}{k_B T} \left\langle \frac{dQ}{dt}      \right\rangle   = - \frac{1}{k_B} \left\langle \frac{dS_{irr}}{dt}   
          \right\rangle <  0. \label{ne} \\ \nonumber
\end{eqnarray}
Here, $d$ is the dimension of the physical space, and $d \sum_i s_i $ becomes  the number of thermostatted degrees of freedom.
$D$ is the phase space dimension, and $dQ/dt$ is the rate with which 
heat is transferred to the bath and gives rise to a positive rate of entropy production, $ \left\langle dS_{irr} / dt \right\rangle$.  
$T$ is the kinetic temperature enforced by the thermostat. 
These relations show that an infinitesimal phase volume shrinks with an averaged rate, which is 
determined by the rate of heat transfer to the bath, which, in turn, is regulated by the  thermostat variable $\zeta$. 
As a consequence, there exists a multifractal attractor in phase space, to which all trajectories converge for
$t \to \infty$,  and on which the natural measure resides. For Axiom A flows or even non-uniform hyperbolic systems,
these states have been identified with the celebrated Sinai-Ruelle-Bowen (SRB) states \cite{Ruelle:1999}.

This result is quite general  for dynamically thermostatted systems and has been numerically confirmed for many transport processes 
in many-body systems. The computation of the Lyapunov spectrum is essential for the understanding of this 
mechanism  \cite{HHP_1987,PH88,PH_1989} and, still, is the only practical way to compute the information dimension 
of the underlying attractor.
Such a system is  special in the sense that its stationary measure is 
singular and resides on a fractal attractor with a vanishing phase space volume. The ensuing Gibbs entropy is diverging to  minus infinity,
which explains the paradoxical situation that heat is  continuously transferred to the bath giving rise to a positive rate of
irreversible entropy production $\dot{S}_{irr}$. The fact that there is a preferred direction for the heat flow  is in accordance with  
the Second Law of thermodynamics, which these systems clearly obey.  The divergence of the Gibbs entropy also indicates that the
{\em thermodynamic entropy}, namely  the derivative of the internal energy with respect to the entropy, is not defined
for such stationary non-equilibrium states. 

\begin{figure}[t]
\centering
\includegraphics[width=0.45\textwidth]{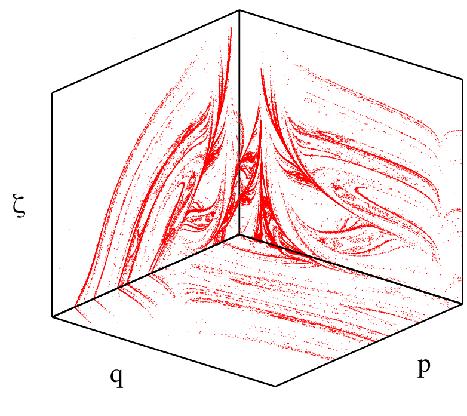}
\caption{(Color online) One-dimensional Frenkel-Kontorova conductivity model described in the main text.
Three Poincar\'e maps for the planes defining  the first octant are shown. The system is in a stationary
non-equilibrium state. Reprinted from Ref.~\cite{Radons_Just}}
\label{frenkel}
\end{figure}
We demonstrate such a fractal attractor for  a one-dimensional Frenkel-Kontorova conductivity model of a charged
particle in a sinusoidal potential, which is subjected to a constant applied field $X$ and a Nos\'e-Hoover thermostat
\cite{HHP_1987,PHH_1990}. The equations of motion are
\begin{equation}
\dot{q} = p; \;\;\; \dot{p} = -\sin(q) + X - \zeta p; \;\;\; \dot{\zeta} = p^2 -1. 
\nonumber
\end{equation} 
The phase space is three-dimensional. In Fig.~\ref{frenkel}  three Poincar\'e maps for mutually orthogonal planes
defining the first octant of the phase space are shown.  The singularity spectrum for this model has been computed in Ref.~\cite{PHH_1990},
and the  multifractal nature of the  stationary phase-space measure has been established.

\begin{figure}[b]
\centering
\includegraphics[width=0.47\textwidth]{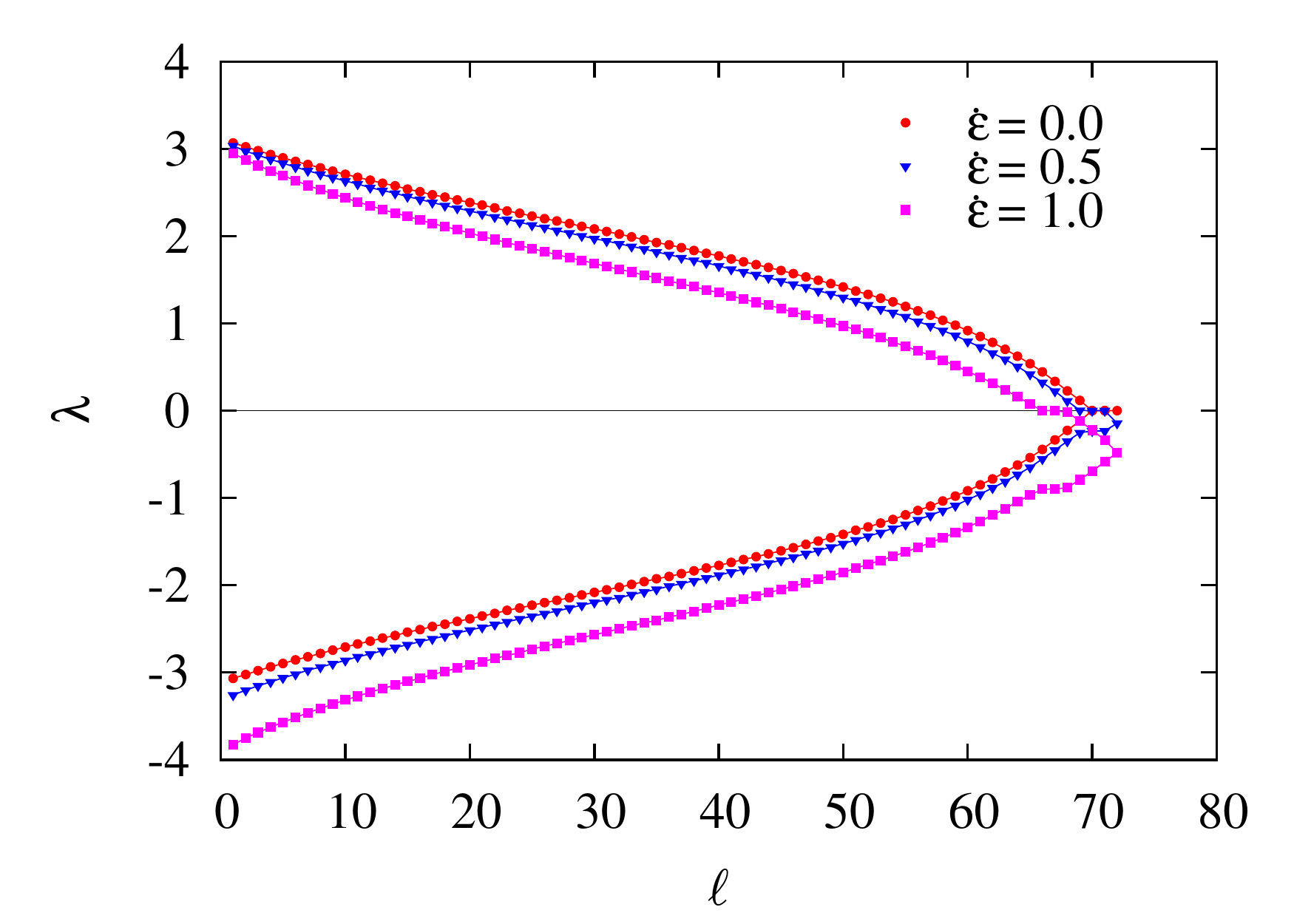}
\caption{(Color online)  Lyapunov spectra of a system of 36 hard disks subjected to iso-energetic planar shear flow with shear rates
$\dot{\varepsilon}$ as indicated by the points.
For details we refer to Ref.~\cite{DPH_1996}.   }
\label{sllod}
\end{figure}

The sum of all Lyapunov exponents becomes negative for stationary non-equilibrium states. 
To demonstrate this effect, we show in Fig.~\ref{sllod} 
the Lyapunov spectra of a system of 36 particles subjected to planar shear flow for various shear rates
$\dot{\varepsilon}$ as indicated by the labels. Homogeneous SLLOD mechanics \cite{EM_2008} in combination with
a Gaussian energy control has been used,  which keeps the internal energy of the fluid a constant of the motion.
The figure emphasizes  the conjugate pairing symmetry in the non equilibrium stationary state:
Since all particles are homogeneously affected by the same thermostat, the pair sum of conjugate exponents
becomes negative and is the same for all pairs. With the Kaplan-Yorke formula, the information dimension is found to be smaller 
than the dimension of the phase space \cite{DPH_1996}, a clear indication of a multi-fractal  phase-space distribution.

\section{Final remarks}
\label{outlook}
Here we have described some of our studies of simple fluids, the results so far obtained, and the new questions 
they raise. Most of  our  simulations are for planar systems due to the numerical effort required for the computation
of Lyapunov spectra, but the results are expected to carry over to three-dimensional systems. 

What have these studies taught us about fluids? For equilibrium systems, the results about the density dependence 
of the  maximum  Lyapunov exponents and of the mixing time in Sec.~\ref{hard_wca} clearly provide deeper and 
quantitative insight into the foundation and limitations of more familiar models for  dense fluids. Still, the behavior of the exponents 
near phase transitions has not been satisfactorily exploited yet and should be  the topic for further research.
It is interesting to note that time-dependent local Lyapunov exponents have been used to identify atypical trajectories 
with very high or very low chaoticity.  Such trajectories, although rare and difficult to locate in phase space, may be of 
great  physical significance.  For example, in phase traditions, or in chemical reactions,  they may lead over unstable saddle points  
connecting a quasi-stable state, or chemical species, with another. Tailleur and Kurchan invented a very promising 
algorithm, {\em Lyapunov weighted dynamics}, for that purpose \cite{Kurchan}, in which  a swarm of phase points 
evolves according to the natural dynamics. Periodically, trajectories are killed, or cloned,  with a rate determined by their
local exponents $\Lambda_1$.  As a consequence, trajectories are weighted according to their chaoticity.
A slightly modified version of this algorithm, {\em Lyapunov weighted path sampling}, has been suggested by 
Geiger and Dellago \cite{Geiger}, and has been successfully applied to the isomerizing dynamics 
 of a double-well dimer in a solvent.
 
Since the turn of the century,  Lyapunov modes have raised a lot of interest. But it has been surprisingly difficult 
to connect them to other hydrodynamic properties \cite{Zabey,CTM_2011,MT_2013}. Most recently, however,  some progress 
has been made. But this problem will still occupy researchers for some time. 

Another interesting and promising field is the study of systems with qualitatively different degrees of freedom, such as
translation and rotation. Our studies of  the Lyapunov instability of rough disks and of linear dumbbell molecules has opened another
road to investigate translation-rotation coupling. An extension of these studies to vibrating molecules and
to more complex systems  with  internal rotation, such as rotational isomer dynamics in butane  \cite{PT_unpublished}, 
are promising topics for future research. 

Arguably, the most important impact Lyapunov spectra have achieved is in non-equilibrium  statistical mechanics. In combination with 
dynamical thermostats, it is possible to generate  stationary states far from equilibrium and to link the rate of entropy production with the
time-averaged friction coefficients and the logarithmic rate of phase-volume contraction. See Eq.(\ref{ne}). The existence of a multi-fractal
phase-space distribution with an information dimension smaller than the phase space dimension provides a geometrical
interpretation of the Second Law of thermodynamics. The importance of this result is reflected in the voluminous literature on this
subject.  
 
 There are many other applications of local Lyapunov exponents and tangent vectors in meteorology, in 
 the geological science, and in other fields.
 For more details we refer to Ref.~\cite{Cencini}, which is a recent collection of 
 review articles and applications.

\section{Acknowledgments}

We thank Prof. Christoph Dellago and  Prof. Willliam G. Hoover for many illuminating discussions.
We  also  thank  Drs. Ljubo Milanovi\'c,   Jacobus van Meel,   Robin Hirschl and 
Christina Forster, whose enthusiasm and dedication provided the basis for the present discussion.
One of us (HAP) is also grateful to  the organizers of a workshop at KITPC in July 2013,  whose  invitation
provided the possibility to discuss various aspects of this work with other participants of that event.


\begin{references}

\bibitem{MPH_1998}  Lj. Milanovi\'c, H.A. Posch, and Wm.G. Hoover, Molec. Phys., {\bf 95}, 281 - 287 (1998).
\bibitem{PH88} H.A. Posch and Wm.G. Hoover, Phys. Rev. A {\bf 38}, 473 (1988).
\bibitem{PH_1989} H.A. Posch, and W.G. Hoover, Phys. Rev. A {\bf 39}, 2175  (1989). 
\bibitem{Gray} C.G. Gray and K.E. Gubbins, {\em Theory of molecular fluids}, Oxford University Press, Oxford (1984). 
\bibitem{Alder} B.J. Alder and T.E. Wainwright, J. Chem. Phys. {\bf 27}, 1208 (1957).
\bibitem{DPH_1996} C. Dellago, H.A. Posch and Wm.G. Hoover, Phys. Rev. E {\bf 53}, 1485 (1996).
\bibitem{Allen} M.P. Allen and D.J. Tildesley, {\em Computer Simulation of Liquids}, Clarendon Press, Oxford (1987).    
\bibitem{chapman:1953} S. Chapman and T.G. Cowling, {\em The mathematical theory of 
       non-uniform gases}, 3rd. edition, Cambridge University Press, Cambridge (1990). 
\bibitem{AI_1987} M.P. Allen and A.A. Imberski, Molec. Phys. {\bf 60}, 453 (1987).
\bibitem{TS_1980} D.J. Tildesley and W.B. Streett, Molec. Phys. {\bf 41} 85 (1980).
\bibitem{VB} J. Vieillard-Baron, Molec. Phys. {\bf 28}, 809 (1974).
\bibitem{RS} D.W. Rebertus and K.M. Sando, J. Chem. Phys.  {\bf 67}, 2585 (1977).    
\bibitem{FMM} D. Frenkel, B.M. Mulder, and J.P. McTague, Phys. Rev. Lett. {\bf 52}, 287 (1984).
\bibitem{TAEFK} J. Talbot, M.P.Allen, G.T. Evans, D. Frenkel, and D. Kivelson, Phys. Rev. A {\bf 39}, 4330 (1989).  
\bibitem{FMag} D. Frenkel and J.F. Maguire, Molec. Phys. {\bf 49}, 503 (1983).    
\bibitem{BP_2013} H. Bosetti and H.A. Posch, J. Phys. A:  Math. Theor. {\bf 46}, 254011 (2013).
\bibitem{Bdiss} H. Bosetti, {\em On the microscopic dynamics of particle systems in and out of
            thermal equilibrium}, Ph.D. thesis, University of Vienn, (2011).
\bibitem{Milano} Lj. Milanovi\'c and H.A. Posch, J. Molec. Liquids, {\bf 26-27}, 221 (2002).
\bibitem{MPH_chaos} Lj. Milanovi\'c, H.A. Posch, and Wm.G. Hoover, Chaos, {\bf 8}, 455 - 461 (1998).
\bibitem{Oseledec:1968} V.I. Oseledec, Trudy Moskow. Mat. Obshch. {\bf 19}, 179, (1968) 
                         [Trans. Mosc. Math. Soc. {\bf 19}, 197 (1968).
\bibitem{Ruelle:1979} D. Ruelle,  Publications Math\'ematiques de l'IH\'ES {\bf 50}, 27 (1979).
\bibitem{Eckmann:1985} J.-P. Eckmann and D. Ruelle, Rev. Mod. Phys. {\bf 57}, 617 (1985).
\bibitem{Ruelle:1999} D. Ruelle, J. of Statist. Phys. {\bf 85}, 393 (1999).
\bibitem{Benettin} G. Benettin, L. Galgani, A. Giorgilli, and J.-M. Strelcyn,
              Meccanica {\bf 15}, 21 (1980).
\bibitem{Shimada} I. Shimada and T. Nagashima, Prog. Theor. Phys. {\bf 61}, 1605 (1979).
\bibitem{Wolf} A. Wolf, J.B. Swift, H.L. Swinney, and J.A. Vastano, Physica D {\bf 16}, 285 (1985). 
\bibitem{recipes} W.H. Press, S.A. Teukolsky, W.T. Vetterling, and B.P. Flannery,
       {\em Numerical Recipes in Fortran 77: The Art of Scientific Computing}, 
       Cambridge University Press, Cambridge (1999).                
\bibitem{HP1987} W.G. Hoover, H.A. Posch, and S. Bestiale,  J. Chem. Phys. {\bf 87}, 6665 (1987).
\bibitem{Goldhirsch} I. Goldhirsch, P.-L. Sulem, and S.A. Orszag, Physica D {\bf  27}, 311 (1987).
\bibitem{PH04} H.A. Posch and Wm.G. Hoover, Physica D {\bf 187}, 281 (2004).
\bibitem{BPDH} H. Bosetti, H.A. Posch, C. Dellago, and Wm.G. Hoover, Phys.Rev. E {\bf 82}, 046218 (2010).
\bibitem{Ershov} S.V. Ershov and A.B. Potapov, Physica D {\bf 118}, 167 (1998).
\bibitem{Henon}  M. H\'enon, Comm. Mathem. Phys. {\bf 50}, 69 (1976).
\bibitem{Meyer} H.-D. Meyer, J. Chem. Phys. {\bf 84}, 3147 (1986).
\bibitem{Ginelli} F. Ginelli, P. Poggi, A. Turchi, H. Chat\'e, R. Livi, and A. Politi,  
                        Phys. Rev. Lett. {\bf 99}, 130601 (2007). 
\bibitem{Ginelli_2013} F. Ginelli, H. Chat\'e, R. Livi, and A. Politi,  J. Phys. A: Math. Theor. {\bf 46},  254005  (2013).
\bibitem{BP_2010} H. Bosetti and H.A. Posch, Chem. Physics {\bf 375}, 296 (2010).
\bibitem{Wolfe} C.L. Wolfe and R.M. Samelson, Tellus {\bf 59}A, 355 (2007).
\bibitem{Romero} M. Romero-Bastida, D. Paz\'o, J.M. L\'opez, and M.A.  Rodr\'iguez,  Phys. Rev. E {\bf 82}, 036205 (2010).
\bibitem{WHH} F. Waldner, Wm.G. Hoover, and C.G. Hoover, Chaos, Solitons and Fractals {\bf 60}, 68 (2014).
\bibitem{FP_2005} Ch. Forster and H.A. Posch, New Journal of Physics, {\bf 7}, 32 (2005).
\bibitem{WCA1} H.C. Anderson, D. Chandler, and J.D. Weeks, Adv. Chem. Phys. {\bf 34}, 105 (1976).
\bibitem{WCA2} D. Chandler, J.D. Weeks, and H.C. Anderson, Science {\bf 220}, 787 (1983). 
\bibitem{HBP_1998} Wm.G. Hoover, K. Boerker, and H.A. Posch, Phys. Rev. E {\bf 57}, 3911 (1998).
\bibitem{FHPH_2004} Ch. Forster, R. Hirschl, H.A. Posch, and Wm.G. Hoover, Physica D {\bf 187}, 294 (2004).
\bibitem{Pikovsky_1} A. Pikovsky and A. Politi, Nonlinearity {\bf 11}, 1049 (1998).
\bibitem{Pikovsky_2} A. Pikovsky and A. Politi, Phys. Rev. E {\bf 63}, 036207 (2001).
\bibitem{TM_2003a} T. Taniguchi and G.P. Morriss, Phys. Rev. E {\bf 68}, 026218 (2003).
\bibitem{TM_2003b} T. Taniguchi and G.P. Morriss, Phys. Rev. E {\bf 68}, 046203 (2003).
\bibitem{Pesin} Ya.B. Pesin, Usp. Mat. Nauk {\bf 32}, 55 (1977); Russ. Math. Survey {\bf 32} 55 (1977).
\bibitem{Arnold} V.I. Arnold and A. Avez, {\em Ergodic Problems of Classical Mechanics}, Addison Wesley Publishing Company, New York/Amsterdam (1968).
\bibitem{Zaslav} G.M. Zaslavsky, {\em Hamiltonian Chaos and Fractional Dynamics}, Oxford University Press, New York (2005).  
\bibitem{DP_relax} C. Dellago and H.A. Posch, Phys. Rev. E {\bf 55}, R9 (1997).
\bibitem{Zon} R. van Zon and H. van Beijeren,  J. Stat. Phys. {\bf 109}, 641 (2002).
\bibitem{Wijn} A.S. de Wijn, Phys. Rev. E {\bf 71}, 046211 (2005).
\bibitem{BDPD} H. van Beijeren,  J.R.  Dorfman, H.A. Posch, and C. Dellago, Phys. Rev. E {\bf 56}, 5272 (1997).
\bibitem{ZBD} R. van Zon, H. van Beijeren, and C. Dellago, Phys. Rev. Lett {\bf 80}, 2053 (1998).
\bibitem{Stillinger} T.M. Truskett, S. Torquato, S. Sastry,
             P. G. Debenedetti, and F.H. Stillinger, Phys. Rev. E, {\bf 58}, 3083 (1998).
\bibitem{Tox} S. Toxvaerd, Phys. Rev. Lett. {\bf 51}, 197 (1983). 
              Please note that the density used by S. Toxvaerd needs to be divided by $2^{1/3}$ to conform 
              to the units used by us.       
\bibitem{PHH_1990} H.A. Posch, Wm.G. Hoover, and B.L. Holian, Ber. Bunsenges. Phys. Chem. {\bf 94}, 250 (1990). 
\bibitem{DP_3d} C. Dellago and H.A. Posch, Physica A {\bf 240}, 68 (1997).
\bibitem{Szasz_Buch} H.A. Posch and R. Hirschl, ``Simulation of Billiards and of 
      Hard-Body Fluids'', pages 269 - 310, in {\em Hard Ball Systems and the Lorenz Gas}, 
      edited by D. Szasz, Encyclopedia of the mathematical sciences
      {\bf 101}, Springer, Berlin (2000).
\bibitem{Hthesis} R. Hirschl, {\em Computer simulation of hard-disk and hard-sphere systems: ''Lyapunov modes'' and
              stochastic color conductivity}, Master thesis, University of Vienna,  1999. 
\bibitem{Eck} J.-P. Eckmann and O. Gat, J. Stat. Phys. {\bf 98}, 775 (2000). 
\bibitem{McNamara} S. McNamara and M. Mareschal, Phys. Rev. E {\bf 64}, 051103 (2001).
\bibitem{Mare} M. Mareschal and S. McNamara, Physica D, {\bf 187}, 311 (2004).    
\bibitem{Goldstone} J. Goldstone,  Nuovo Cimento {\bf 19},  154 (1961).
\bibitem{Forster} D. Forster, {\em Hydrodynamic Fluctuations, Broken Symmetry, and Correlation Functions}, The Benjamin/Cummings Publishing Company, Reading (1975).        
\bibitem{Wijn_vB} A.S. de Wijn and H. van Beijeren, Phys. Rev. E {\bf 70}, 016207 (2004).  
\bibitem{TDM} T. Taniguchi, C.P. Dettmann, and G.P. Morriss, J. Stat. Phys. {\bf 109}, 747 (2002).
\bibitem{TM_2002} T. Taniguchi and G.P.  Morriss, Phys. Rev. E {\bf 65}, 056202 (2002). 
\bibitem{HPFDZ} Wm.G. Hoover, H.A. Posch, Ch. Forster, Ch. Dellago, and M. Zhou, J. Stat. Phys. {\bf 109}, 765 (2002).   
\bibitem{Zabey} J.-P- Eckmann, Ch. Forster, H.A. Posch, and E. Zabey,  J. Stat. Phys. {\bf 118}, 813 - 847 (2005).  
\bibitem{Radons_Yang} G. Radons and H.-L. Yang, {\em Static and dynamic correlations in many-particle Lyapunov vectors}, arXiv:nlin/0404028.     
\bibitem{Yang_Radons} H.-L.Yang and G. Radons, Phys. Rev. E {\bf 71}, 036211 (2005).        
\bibitem{FHP_Congress} Ch. Forster, R. Hirschl, and H.A. Posch, Proceedings of the
         XIVth International Congress on Mathematical Physics, edited by J.-C. Zambrini, World Scientific, Singapore (2003). 
\bibitem{CTM_2011} T. Chung, D. Truant, and G.P. Morriss, Phys. Rev. E  {\bf 83}, 046216 (2011).
\bibitem{CTM_2010} T. Chung, D. Truant, and G.P. Morriss, Phys. Rev. E  {\bf 81}, 066208 (2010).        
\bibitem{MT_2013} G.P. Morriss and D.P. Truant, J. Phys. A:  Math. Theor. {\bf 46}, 254010 (2013).
\bibitem{bryan:1894} G.H. Bryan, Reports of the  British Association for the  Advancement of  Science, Vol. 64, p. 64 - 101
       (p. 83 in particular), (1894).
\bibitem{pidduck:1922} F.B. Pidduck, Proc. R. Soc. A, {\bf 101}, 101 (1922).
\bibitem{Berne_I} J. O'Dell  and B.J. Berne, J. Chem. Phys. {\bf 63}, 2376 (1975).
\bibitem{Berne_II} B.J. Berne, J. Chem. Phys. {\bf 66}, 2821 (1977).      
\bibitem{Berne_III} C.S. Pangali and B.J. Berne, J. Chem. Phys. {\bf 67}, 4561 (1977).
\bibitem{Berne_IV} J.A. Montgomery, Jr. and B.J. Berne, J. Chem. Phys. {\bf 67}, 4580 (1977).
\bibitem{vMP} J. van Meel and H.A. Posch, Phys. Rev. E {\bf 80}, 016206 (2009).
\bibitem{Bellemans_1980} A. Bellemans, J. Orban, and D. Van Belle,  Molec. Phys. {\bf 39}, 781 (1980).
\bibitem{Mdiss} Lj. Milanovi\'c, {\em Dynamical instability of two-dimensional molecular fluids: hard dumbbells},
               Ph.D. thesis,  University of Vienna (2001).
\bibitem{Borzsak} I. Borzs\'ak, H.A. Posch, and A. Baranyai,  Phys. Rev. E {\bf 53}, 3694 (1996).
\bibitem{Kum} O. Kum, Y.H. Shin, and E.K. Lee, Phys. Rev. E {\bf 58}, 7243 (1998).
\bibitem{HHP_1987}  B.L. Holian, W.G. Hoover and H.A. Posch, Phys. Rev. Lett. {\bf 59}, 10 (1987).  
\bibitem{EM_2008} D.J. Evans and G. Morriss, {\em Statistical Mechanics of Nonequilibrium Liquids}, 2nd edition, Cambridge University Press, Cambridge (2008). 
\bibitem{HH_2012} Wm.G. Hoover and C.G. Hoover, {\em Time Reversibility, Computer Simulation, Algorithms, Chaos},
                  2nd edition, World Scientific Publishing Company, Singapore (2012).
\bibitem{Radons_Just} H.A. Posch and Ch. Forster, '' Lyapunov Instability if Fluids'', in {\em Collective Dynamics of
               Nonlinear and Disordered Systems}, G. Radons, W. Just, and P. H\"aussler eds, Springer Verlag, Berlin (2005).
\bibitem{Kurchan} J. Tailleur and J. Kurchan, Nature Phys. {\bf 3}, 203 (2007).
\bibitem{Geiger} P. Geiger and C. Dellago, Chem. Phys. {\bf 375}, 309 (2010). 
\bibitem{PT_unpublished} H.A. Posch and S. Toxvaerd, unpublished.                          
\bibitem{Cencini} M. Cencini and F. Ginelli eds., Special issue on {\em Lyapunov analysis: from dynamical systems theory to applications}, 
               J. Phys. A, {\bf 46} (2013).
               
\end{references}
\end{document}